\documentclass{aa}  

\usepackage{graphicx}
\usepackage{natbib,twoopt,epsfig}
\usepackage[breaklinks=true]{hyperref} 
\bibpunct{(}{)}{;}{a}{}{,} 
\makeatletter
\newcommandtwoopt{\citeads}[3][][]{\href{http://adsabs.harvard.edu/abs/#3}%
{\def\hyper@linkstart##1##2{}%
\let\hyper@linkend\@empty\citealp[#1][#2]{#3}}}
\newcommandtwoopt{\citepads}[3][][]{\href{http://adsabs.harvard.edu/abs/#3}%
{\def\hyper@linkstart##1##2{}%
\let\hyper@linkend\@empty\citep[#1][#2]{#3}}}
\newcommandtwoopt{\citetads}[3][][]{\href{http://adsabs.harvard.edu/abs/#3}%
{\def\hyper@linkstart##1##2{}%
\let\hyper@linkend\@empty\citet[#1][#2]{#3}}}
\newcommandtwoopt{\citeyearads}[3][][]%
{\href{http://adsabs.harvard.edu/abs/#3}
{\def\hyper@linkstart##1##2{}%
\let\hyper@linkend\@empty\citeyear[#1][#2]{#3}}}
\makeatother
\usepackage{txfonts,bm}
\begin{document} 

   \title{Alfv\'en wave phase-mixing in flows:}
   \subtitle{Why 
over-dense, solar coronal, 
open magnetic field structures are cool}
   \author{D. Tsiklauri}
   \institute{School of Physics and Astronomy,
Queen Mary University of London,
Mile End Road, London, E1 4NS,
United Kingdom}

   \date{Received 2 August 2015; accepted 4 November 2015}

 
  \abstract
   {}
   {The motivation for this study is to include the effect of plasma 
flow in Alfv\'en wave (AW) damping via phase mixing and 
to explore the observational implications. }
   {Our magnetohydrodynamic (MHD) 
simulations and analytical calculations
show that, when a background flow is present, 
mathematical expressions for the AW
damping via phase mixing are modified by the
following substitution: $C_A^\prime(x) \to C_A^\prime(x)+V_0^\prime(x)$, 
where $C_A$ and $V_0$ are AW phase 
and the flow speeds, and the prime denotes a derivative 
in the direction across the 
background magnetic field.}
{In uniform magnetic fields and 
over-dense plasma structures, where $C_A$ is smaller
than in the surrounding plasma, the flow, which is confined
to the structure and going
 in the same direction as the AW, reduces the effect of 
phase-mixing, because on the edges of the structure 
$C_A^\prime$ and $V_0^\prime$ have opposite signs. Thus, 
the wave damps 
by means of slower phase-mixing compared to the case without the flow.
This is the result of the co-directional 
flow that reduces the wave
front stretching in the transverse direction. 
Conversely, the
counter-directional flow increases the wave
front stretching in the transverse direction, therefore making
the phase-mixing-induced heating more effective.
Although the result is generic and is applicable
to different laboratory or astrophysical plasma systems,
we apply our findings to addressing the question why over-dense solar coronal open 
magnetic field structures 
(OMFS) are cooler than the background plasma.
Observations show that the over-dense OMFS (e.g. solar coronal 
polar plumes) are cooler 
than surrounding plasma and that, in these structures,
Doppler line-broadening  is consistent with bulk plasma
motions, such as AW.}
   {If over-dense solar coronal OMFS are heated by AW damping via phase-mixing,
we show that, co-directional with AW, plasma flow in them 
reduces the phase-mixing induced-heating, thus providing an explanation
of why they appear cooler than the background.}


   \maketitle

\section{Introduction}

A large amount of work has been dedicated to understand the role 
of Alfv\'en wave (AW) damping in providing heating for 
laboratory and astrophysical plasmas, be it the 
solar corona in 
general, or, more particularly, its open magnetic field structures (OMFS).
Observed AW flux is sufficient to heat the 
corona \citep{2005psci.book.....A}.
However, Spitzer resistivity alone is insufficient to 
dissipate AWs efficiently \citep{2003A&A...400.1051T}.
The phase-mixing of harmonic AWs was proposed as a way to 
alleviate this problem by \citet{1983A&A...117..220H}. 
In phase-mixing, harmonic AW amplitude damps in time as
$B_{AW}(x,t)\propto \exp(-\eta C^\prime_A(x)^2 t^3 k^2/6)$, 
where symbols have their usual meaning 
and $C^\prime_A(x)$ denotes an Alfven speed derivative 
in the density inhomogeneity direction
that runs across the background magnetic field
\citep{1983A&A...117..220H}.
The phase-mixing of AWs that have a Gaussian profile
(as opposed to harmonic)  along the background 
magnetic field has slower, power-law damping,
$B_{AW}\propto t^{-3/2}$, as established by
\citet{2002RSPSA.458.2307W}. 
A mathematically more elegant derivation of 
the latter scaling law has been provided by \citet{2003A&A...400.1051T}.
The exponentially diverging magnetic field lines
provide even faster damping 
$B_{AW}=\exp\left(-A_1\exp(A_2 t)\right)$ 
\citep{1989ApJ...336..442S,2000A&A...354..334D,2007A&A...475.1111S}.
\citet{2000ApJ...533..523M} considered small-amplitude 
AW packets in Wentzel-Kramers-Brillouin (WKB) approximation in 
the Arnold-Beltrami-Childress (ABC) magnetic field. The latter can (for certain values
of physical parameters) have 
exponentially diverging magnetic fields, thus also 
providing a superfast (exponent of exponent) AW dissipation. 
\citet{2014PhPl...21e2902T} studied the dissipation of AW 
in ABC fields using 3D magnetohydrodynamic (MHD) simulations (without a WKB restriction) 
and found that perturbation energy grows in time. 
This was attributed to a new instability, whose growth rate 
appears to be dependent on the value of the resistivity 
and the spatial scale (wavelength) of the AW.
\citet{1998A&A...332..795N} studied the nonlinear 
coupling of MHD waves in a cold, compressible 
plasma with a smoothly inhomogeneous low-speed steady flow 
that was directed along the magnetic field in the phase-mixing context. 
Their main focus, however, was on the wave-mode coupling
rather than a possibility that the flow  reduces the effect of
phase-mixing, which we consider here.

The OMFS in the solar 
corona -- and possibly in coronae of other similar stars
-- come in different forms. 
We refer to the "openness" of the magnetic field in a sense
that the structure must be able to sustain
a background flow. These can be
chromospheric upflows induced by magnetic reconnection
in long coronal loops 
(long enough for the magnetic field to be treated
as uniform, to simplify our model)
or coronal polar plumes.
Generally, 
a distinction is drawn between coronal 
holes \citep{2009soco.book.....G}, 
plumes \citep{1997SoPh..175..393D,1997A&A...318..963D,2007ApJ...658..643R} and
more recently dark jets \citep{2015ApJ...801..124Y}.

The solar coronal holes (CH) are 
regions of low-density and low-temperature 
(compared to the background) plasma, which are believed to be a source for
fast ($\approx 800$ km s$^{-1}$) solar wind,
(see Chapter 4.9 in \citet{2005psci.book.....A}.
CH temperature is typically $0.8-0.9$ MK compared to 
surrounding quiet corona that has a temperature of 
$0.9-1.2$ MK. The boundaries of CH can be clearly seen in soft X-ray 
images, because of the absence of hot $1.2-1.5$ MK
plasma in them, when compared to the background.

The solar coronal polar plumes (CPPs) are radial, thin elongated 
structures that are visible
in white light eclipse
photographs as enhancements of density (3--6
times denser than the background), usually located inside 
coronal holes \citep{1997A&A...318..963D}.
Because they are denser means that the plumes appear brighter than the
surrounding media.
In extreme ultraviolet (EUV) spectroheliograms they appear as
shorter spikes near the polar limb.
Some models and observations suggest that the plume plasma
remains much slower and cooler 
than inter-plume plasma up to $2.0 R_{\sun}$. Values above  these  
plasma parameters start to approach the inter-plume values,
matching them at about $3.0 R_{\sun}$. The flow speed  and
temperature increase of the plasma inside plumes is sometimes 
observed \citep{2007ApJ...658..643R}. The latter work explains
the flow speed  and temperature increase by a possibility of
interaction of the CPP's core with the faster and hotter inter-plume material.
Generally, it is debatable 
whether CPP is the source of fast solar wind (see related 
discussion and references in \citet{1997SoPh..175..393D}).
Ultimately, this is related to the question of whether
CPPs have small dipolar magnetic field patches at their base
or are unipolar. Extensive work by \citet{1997SoPh..175..393D}
suggests that CPP have unipolar magnetic fields.
Their Figure 9, however, shows that, despite the magnetic field
being unipolar, it  is still patchy, or, and this is crucial for our model, 
the  Alfv\'en speed varies across the magnetic 
field, 
giving rise to the phase-mixing of AWs.
Also, on the theoretical side, \citet{1997ApJ...476..357O}
have shown that torsional AWs generate
solitary waves  non-linearly and these  may play a crucial role in fast 
solar wind acceleration, which is a separate issue.

The solar coronal dark jets (DJs) are  relatively new 
features \citep{2015ApJ...801..124Y}. 
The coronal jets have been known for some 
time to be a feature of solar coronal hole
observations obtained in  X-ray or EUV
wavelengths.  \citet{2015ApJ...801..124Y} shows examples of DJs 
that are essentially
invisible in EUV image sequences but have a clear signature in
Dopplergrams derived from an EUV emission line. 
Interestingly,
\citet{2007Sci...318.1580C} provide  evidence for 
Alfv\'en waves in solar X-ray jets.

Chapter 8.2.3 in \citet{2005psci.book.....A} provides
an overview of observations 
of Alfv\'en waves  in OMFS.
Because AW do not perturb density, they can be detected
using spectral observations. The non-thermal broadening of
EUV coronal lines typically shows bulk plasma speeds of 
30 km s$^{-1}$ \citep{1998SoPh..181...91D}.
For a typical AW phase speed of 1000 km s$^{-1}$, this means
that AWs in OMFS have amplitudes of 3\% of the background. 
Therefore AWs in OMFS are weakly non-linear.
Different types of waves are present not only in the
corona. \citet{2012ApJ...757..160J}
use high spatial, spectral, and temporal resolution images, obtained using 
both ground- and space-based instrumentation, to investigate the coupling between 
wave phenomena observed at different heights in the solar atmosphere.

The motivation for this study is to include the effect of plasma 
flow in AW damping via phase-mixing and explore its observational implications. 
Section 2 provides the model and analytical calculations.
Section 3 describes the two-dimensional (2D) MHD simulations
that corroborate the analytical results.
Section 4 lists the main conclusions and outlines
suggestions for future observational validation of the
theory that is formulated in this work.

\begin{figure*}
\sidecaption
\includegraphics[width=12cm]{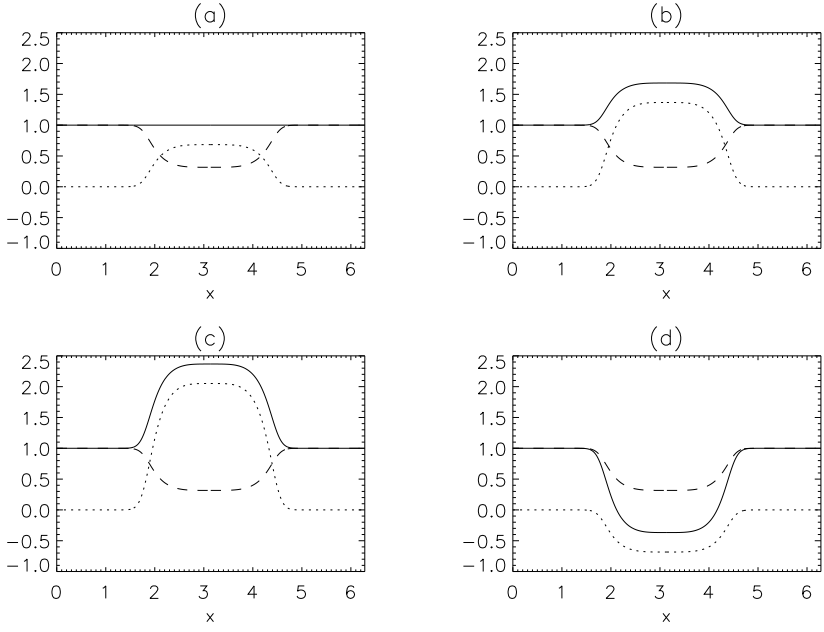}
\caption{Alfv\'en ($C_A(x)$, dashed curve) and background plasma flow 
($V_0(x)$, dotted curve) speeds as a function of $x$-coordinate (across the magnetic field).
Solid curve shows the sum of the two $C_A(x)+V_0(x)$. The different panels
show cases of
(a) $D=1$, flat total speed profile across $x$-coordinate (no phase-mixing),
(b) $D=2$, forward flow exceeding AW speed, 
(c) $D=3$ stronger forward flow, further exceeding the AW speed,
and 
(d) $D=-1$ backward flow.}
\label{fig1}
\end{figure*}

\begin{figure}
\includegraphics[width=\columnwidth]{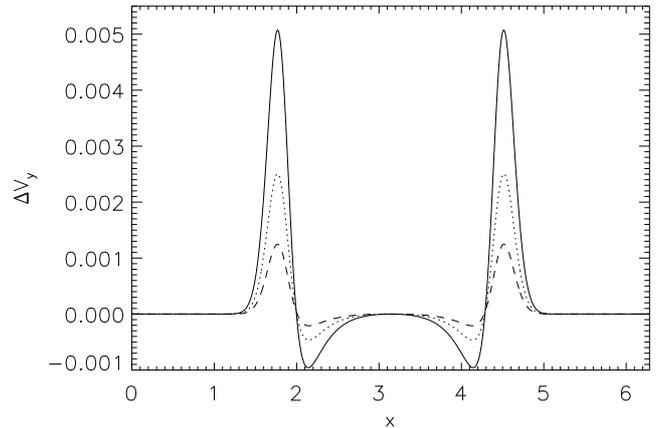}
\caption{Difference between background flow speed at time $t$ as a function of
$x$-coordinate,
$V_0(x,y=y_{max}/2,t)$, and its initial value at $t=0$, $V_0(x,y=y_{max}/2,0)$,
i.e. $\Delta V_y \equiv V_0(x,y=y_{max}/2,t)-V_0(x,y=y_{max}/2,0)$ for different time
instances. Dashed curve corresponds to $t=5$,
 dotted to $t=10$ and solid  to $t=20$. 
This numerical run is considered for the fastest background flow, with $D=3$ 
(as in panel (c) from 
Figure \ref{fig1}).
It is clear that by $t=20$ the flow speed difference is very 
small $\approx 0.005$, i.e. the flow stays intact and does not
disintegrate.}
\label{fig2}
\end{figure}

\begin{figure*}
\centering

\epsfig{file=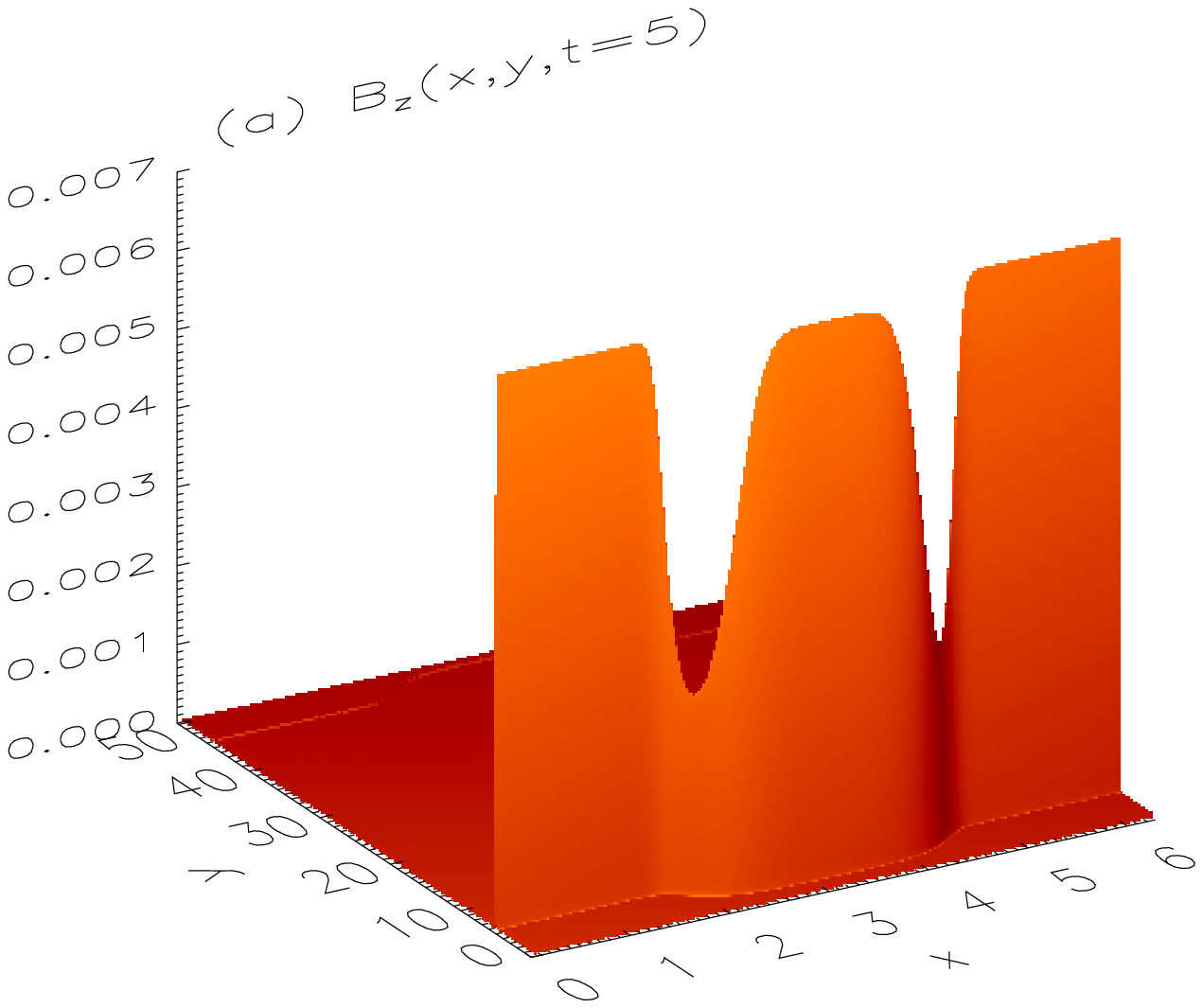,width=8.5cm}
\epsfig{file=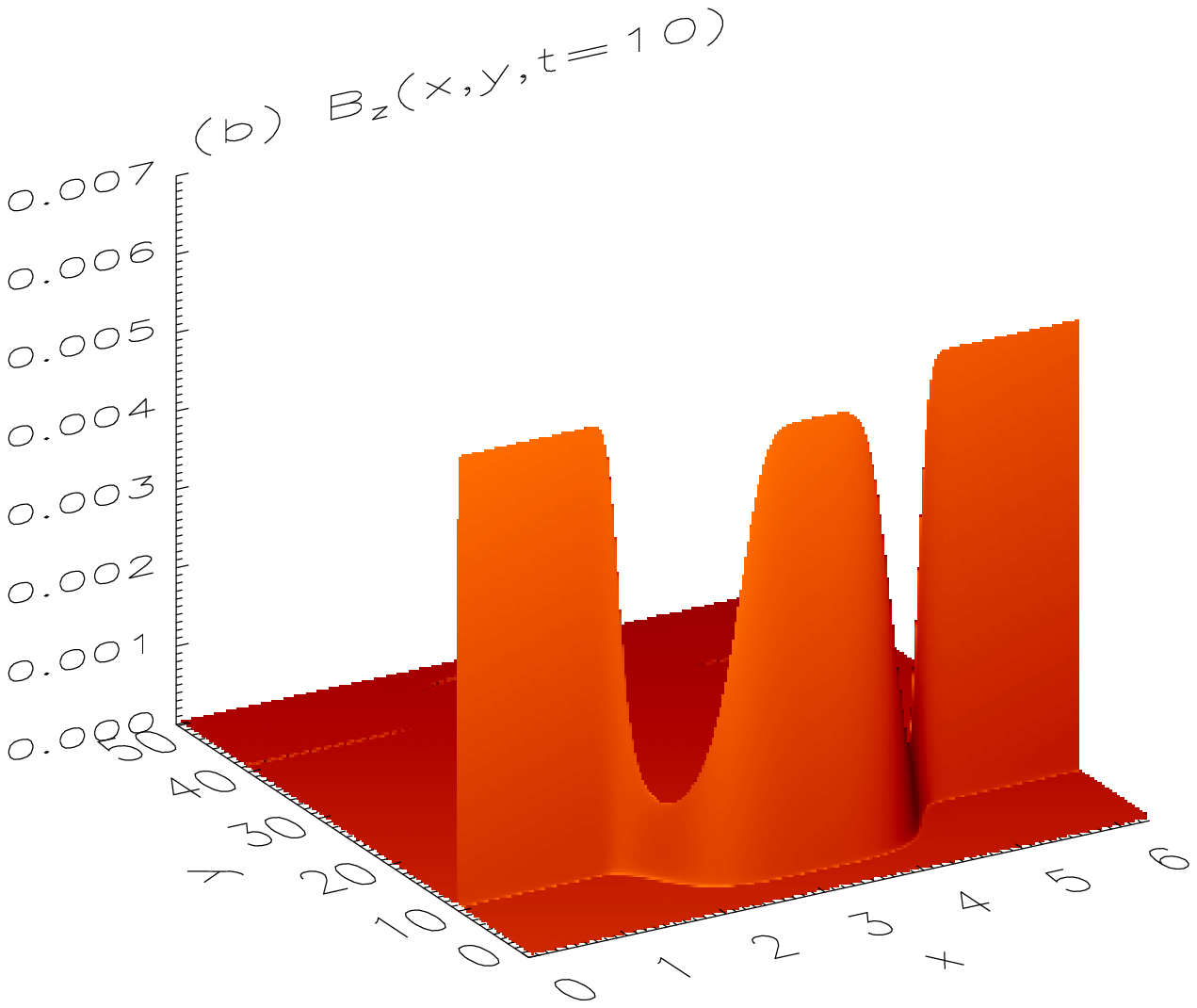,width=8.5cm}
\epsfig{file=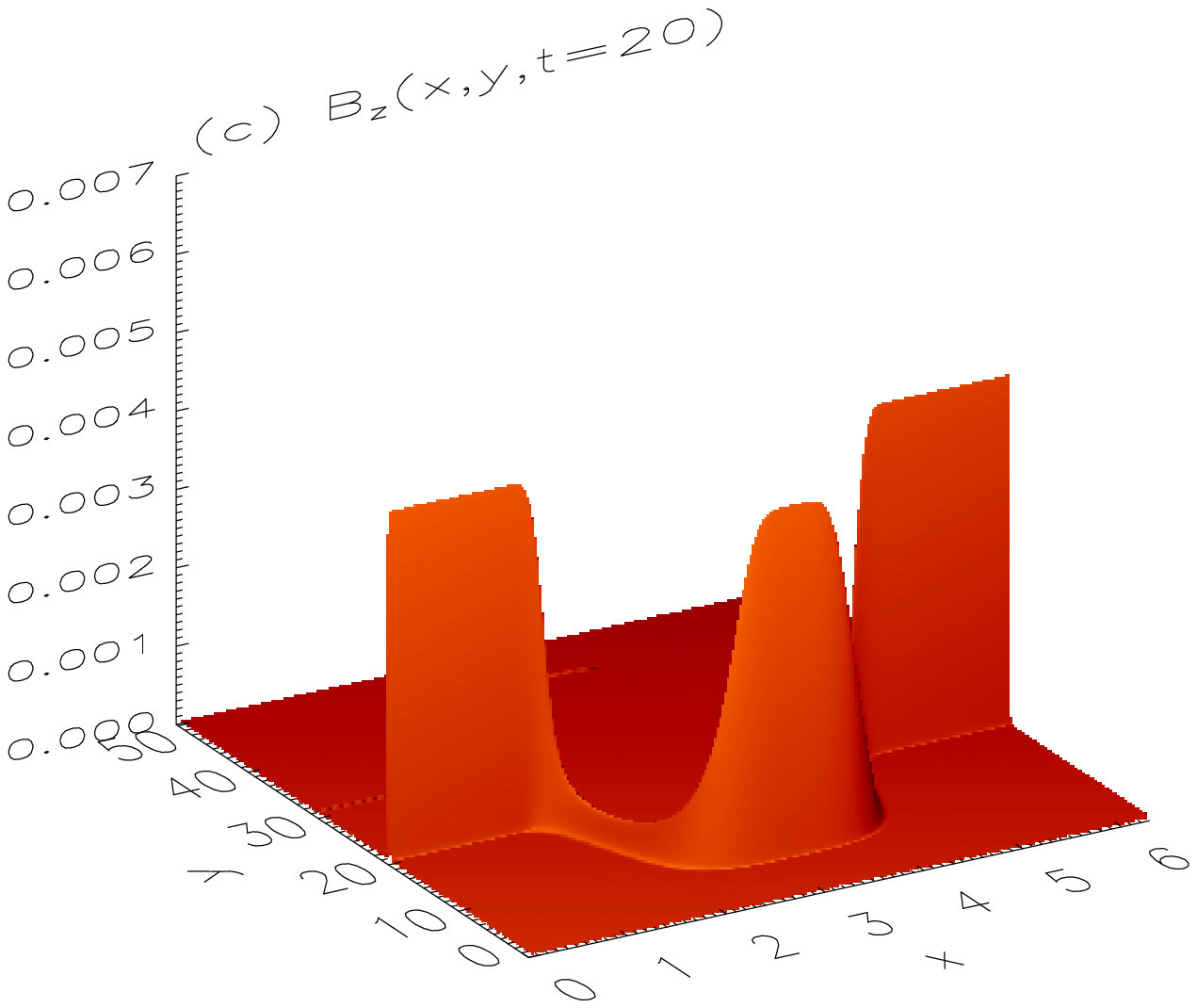,width=8.5cm}
\epsfig{file=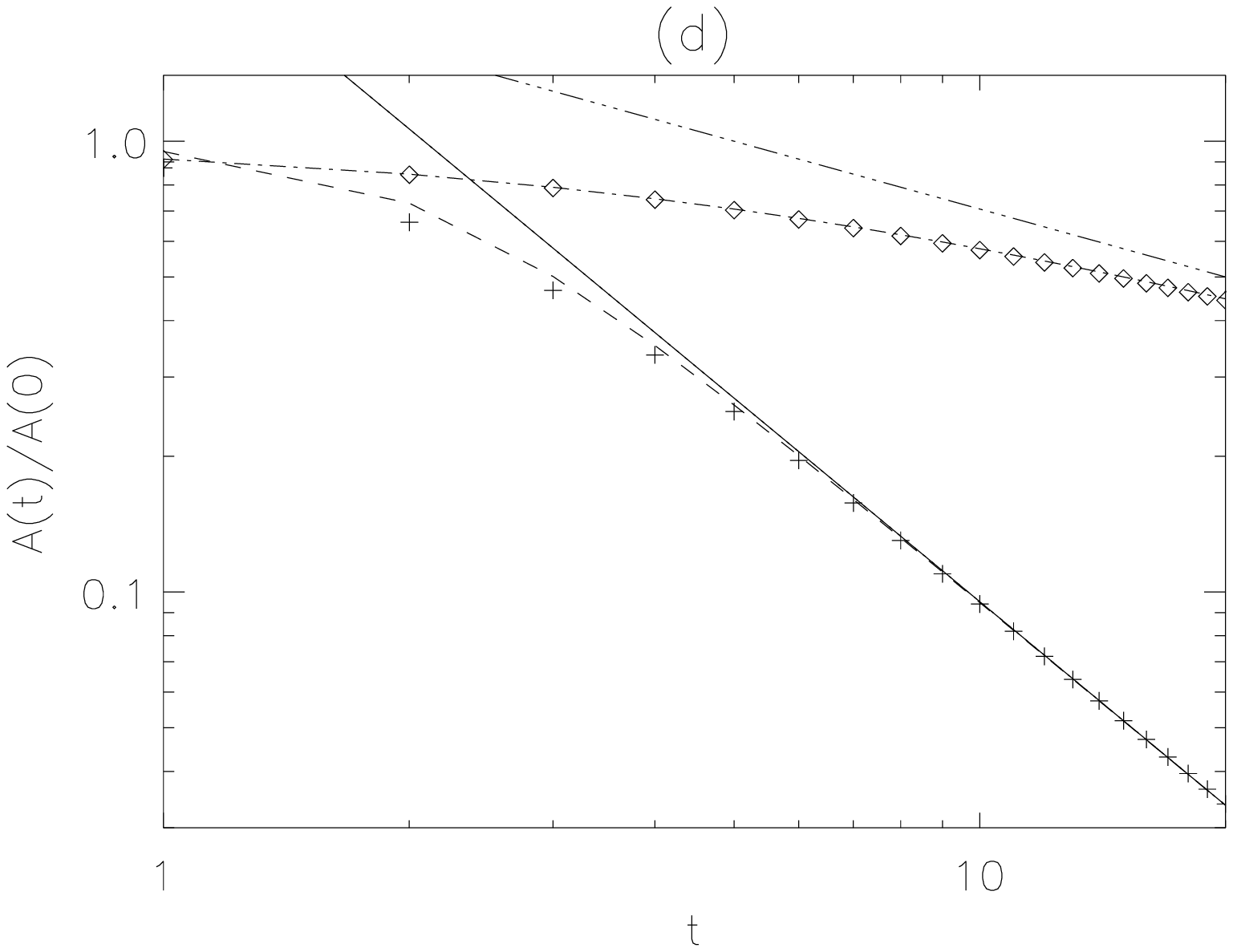,width=8.5cm}
\caption{Shaded surface plots of $B_z(x,y)$ at different times for 
the case without the flow $D=0$. 
Panel (a) is for $t=5$, (b) for $t=10$ and (c) for $t=20$.
Panel (d) is 
time evolution of AW amplitude, normalised to its initial value, for the same case. 
The  solid line corresponds to the asymptotic solution for large times, Eq.(\ref{22}),
at the
strongest density gradient point $x=(907/3000)\times (2\pi)=1.8996$. 
A more general analytical form 
Eq.(\ref{21}) is plotted with dashed curve for the same $x$ value 
(we actually plot $B_z(1.8996,y)/\alpha_0$). 
Crosses and open diamonds are MHD numerical simulation results in the
strongest density gradient point $x=(907/3000)\times (2\pi)=1.8996$ and 
away from the gradient $x=(1/3000)\times (2\pi)=0.0021$ (the first grid 
cell in $x$-direction), respectively, by tracing the maximum value of the 
Gaussian AW pulse.
Dash-triple-dotted line corresponds 
to the asymptotic solution for large times, Eq.(\ref{31}),
which is independent of $x$. 
A more general analytical form 
Eq.(\ref{30}) is plotted with dash-dotted curve. It is also independent of
$x$ because at the peak of the pulse the value of exponent is unity. }
\label{fig3}
\end{figure*}

\begin{figure*}
\centering
\epsfig{file=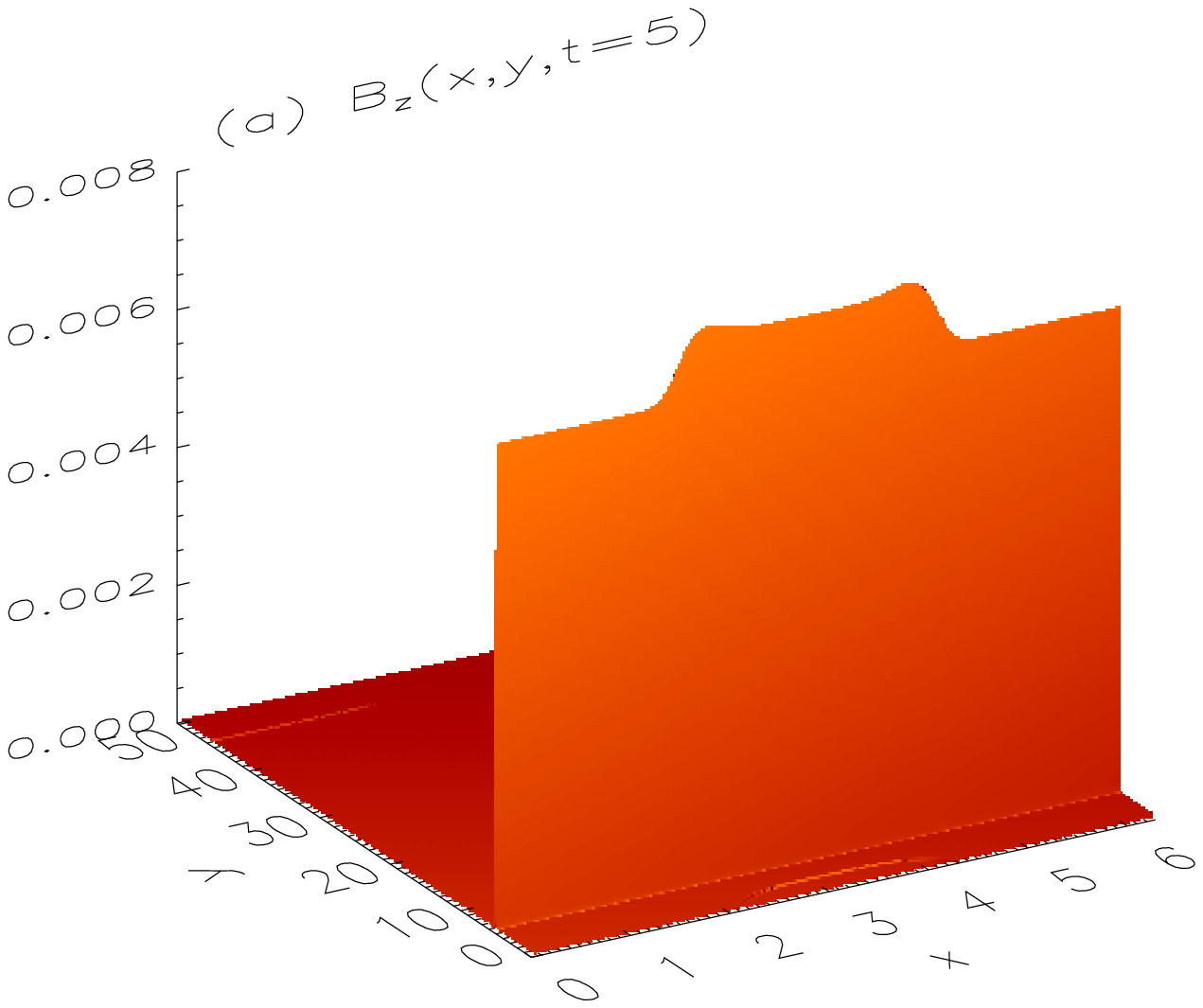,width=8.5cm}
\epsfig{file=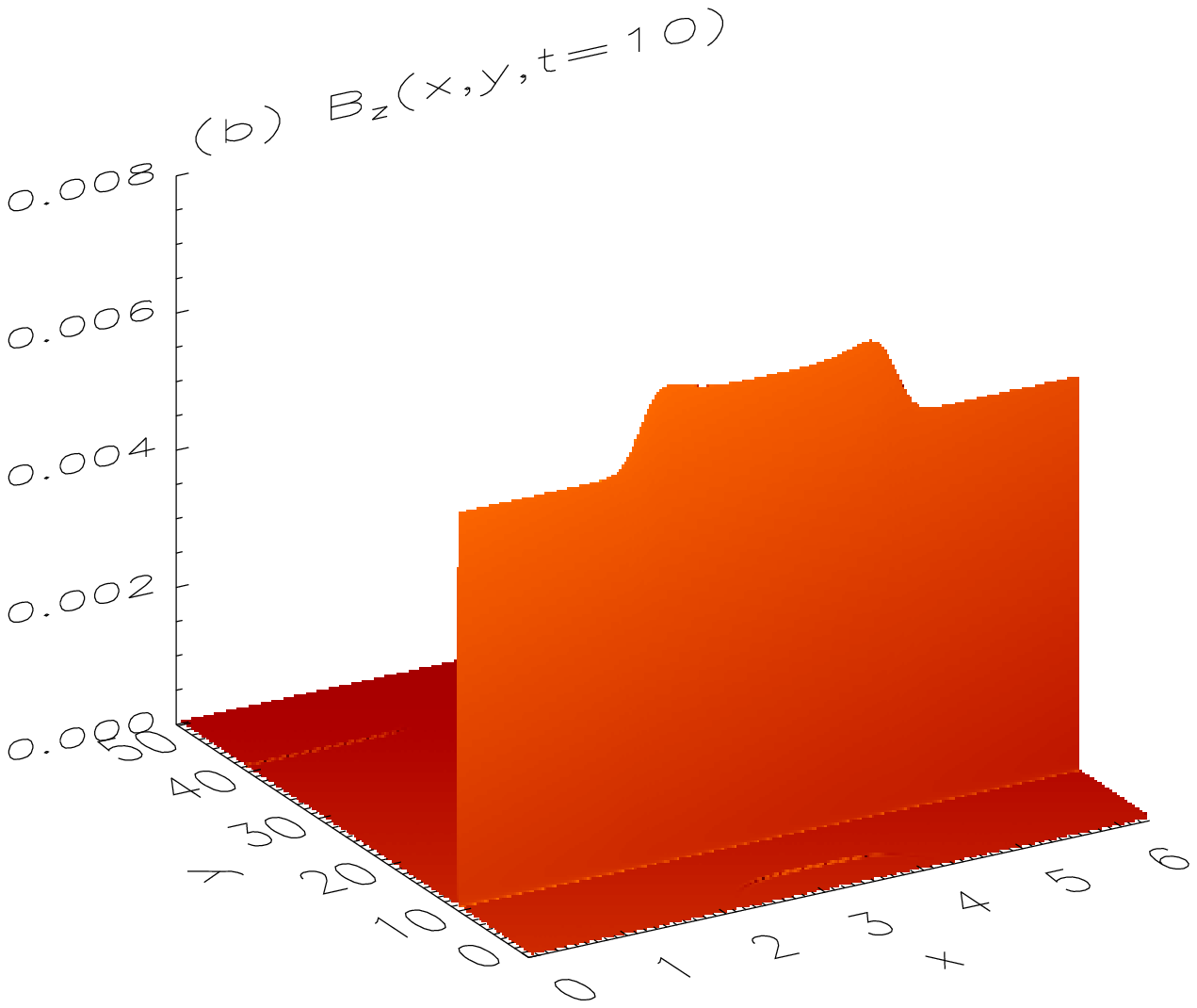,width=8.5cm}
\epsfig{file=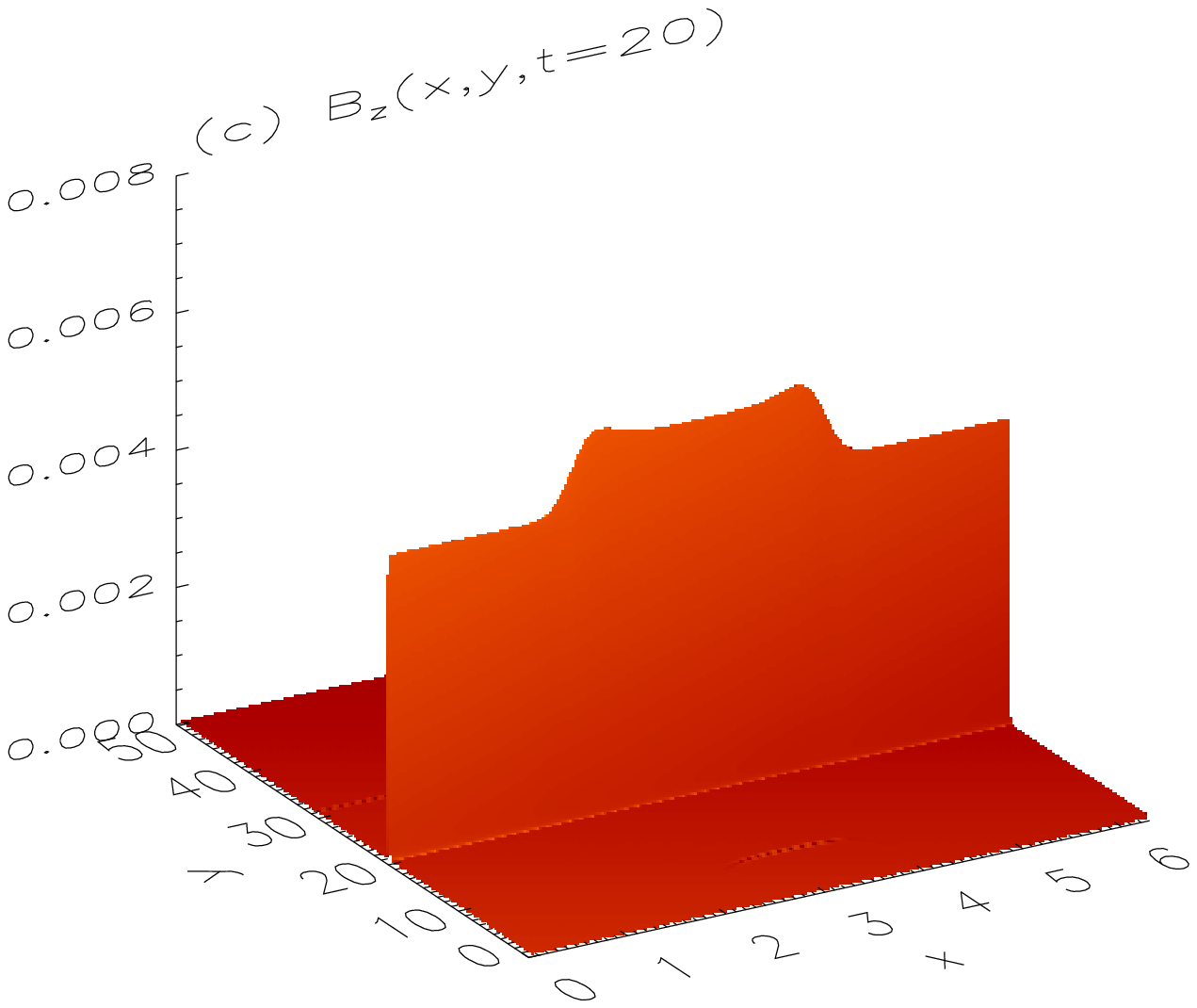,width=8.5cm}
\epsfig{file=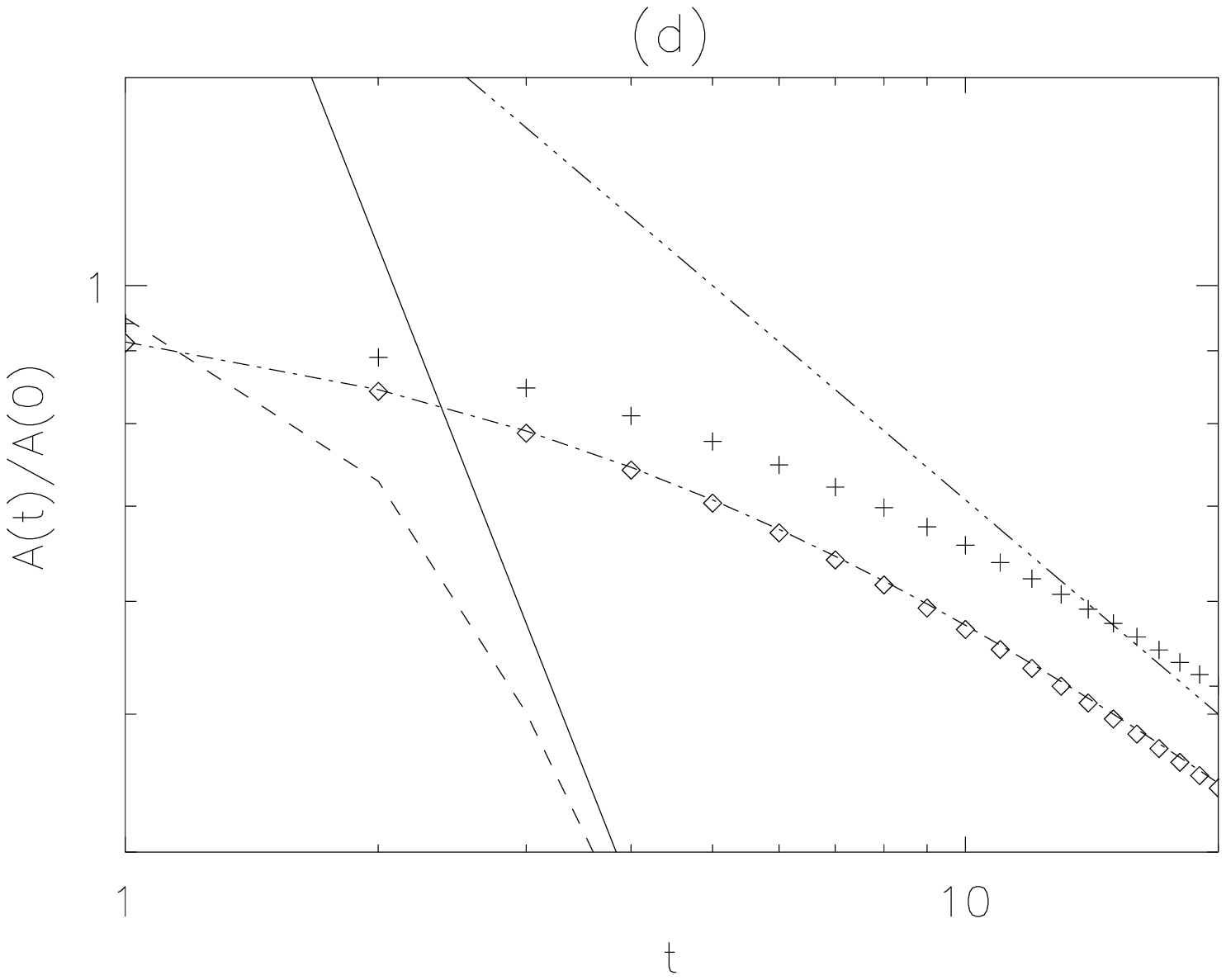,width=8.5cm}
\caption{The same as in Figure \ref{fig3} but for the case of $D=1$.}
\label{fig4}
\end{figure*}

\begin{figure*}
\centering
\epsfig{file=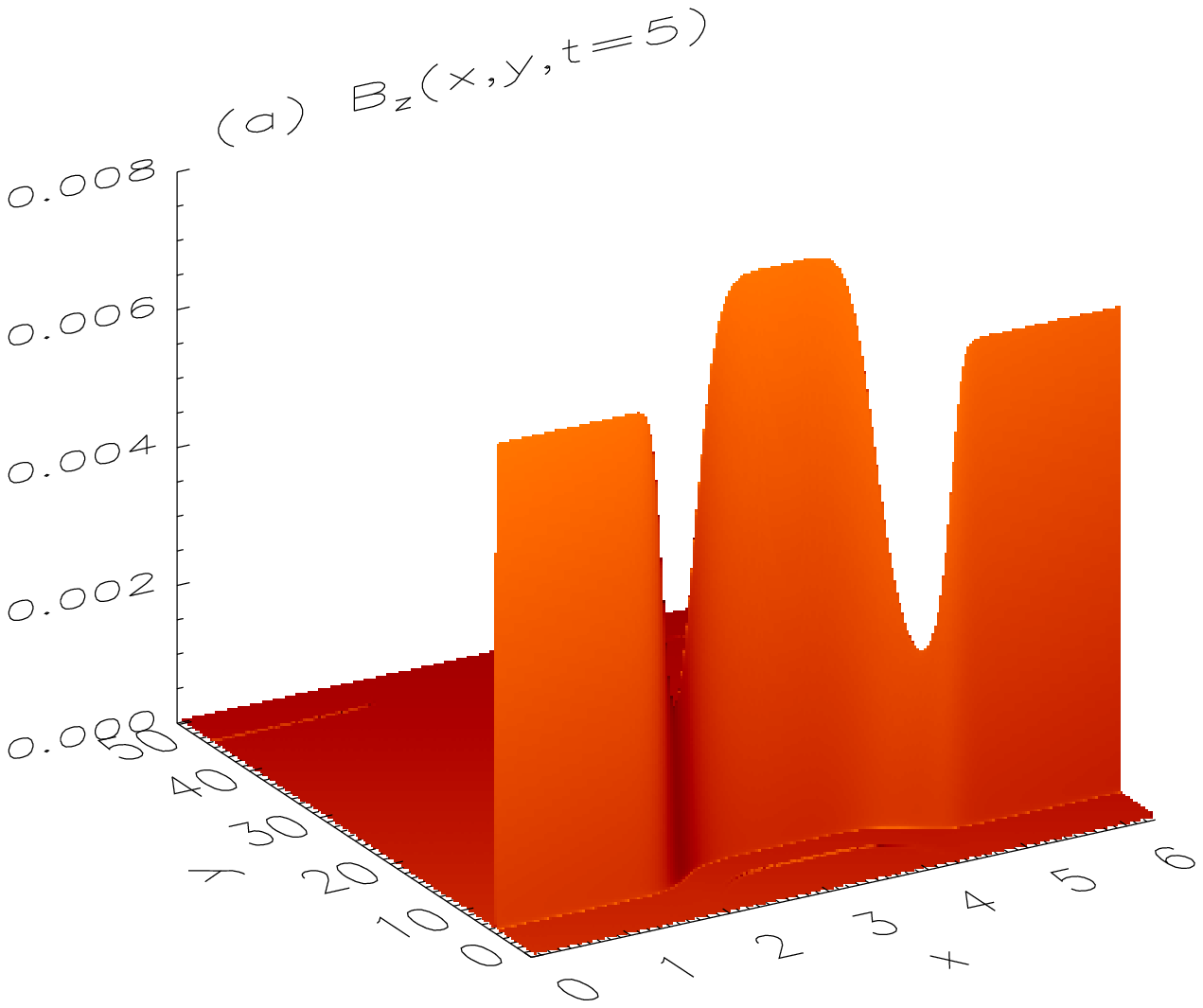,width=8.5cm}
\epsfig{file=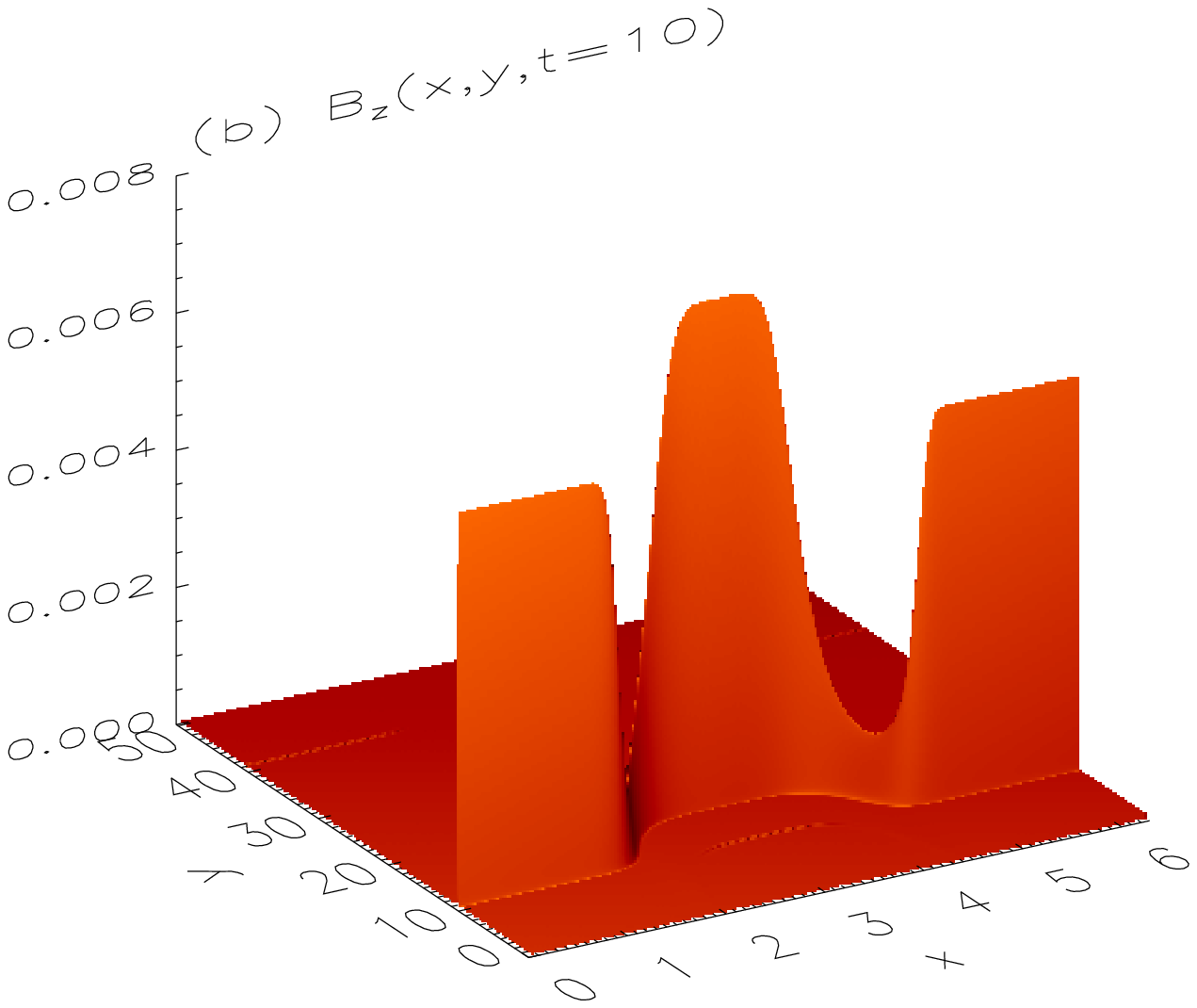,width=8.5cm}
\epsfig{file=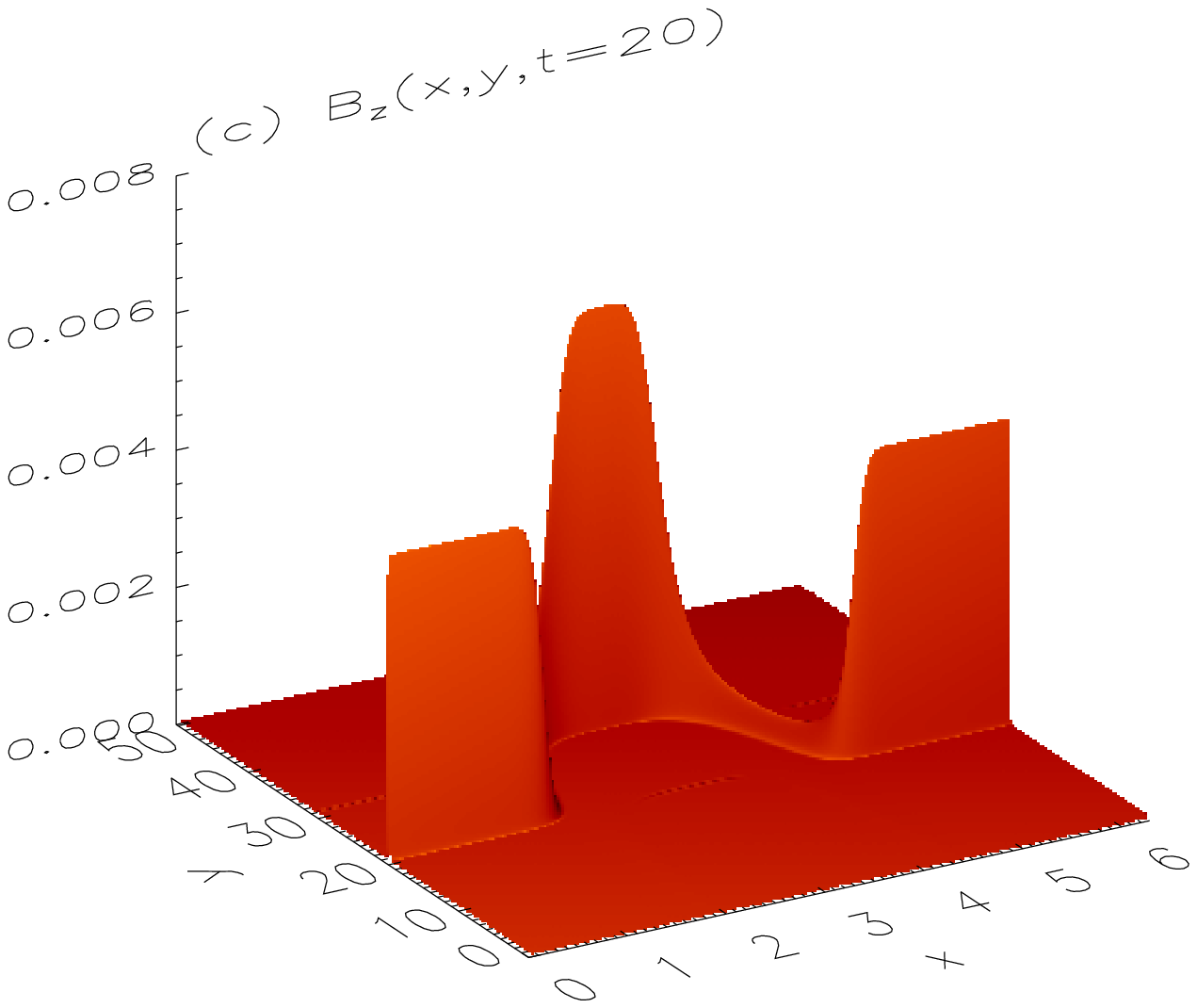,width=8.5cm}
\epsfig{file=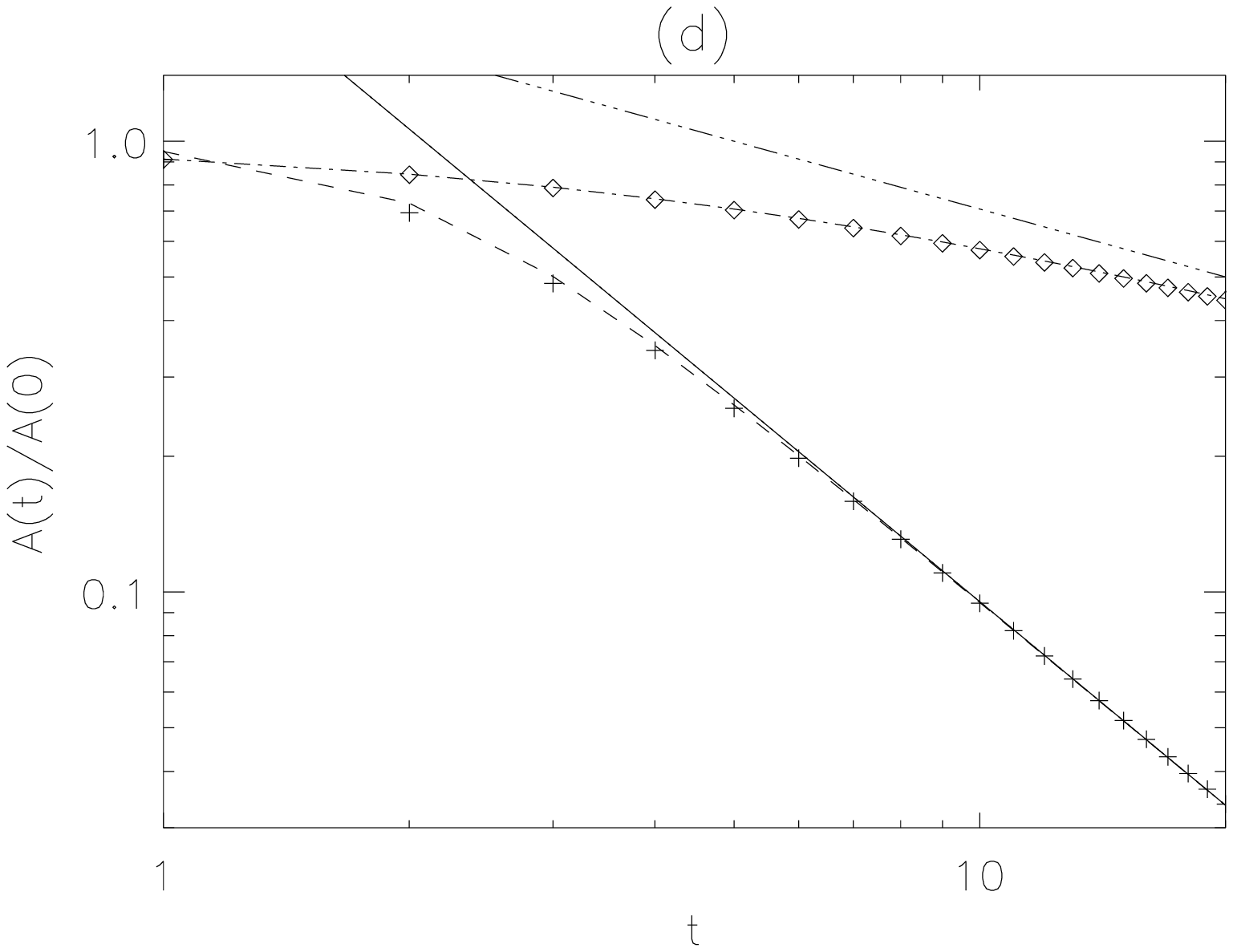,width=8.5cm}
\caption{The same as in Figure \ref{fig3} but for the case of $D=2$.}
\label{fig5}
\end{figure*}

\begin{figure*}
\centering
\epsfig{file=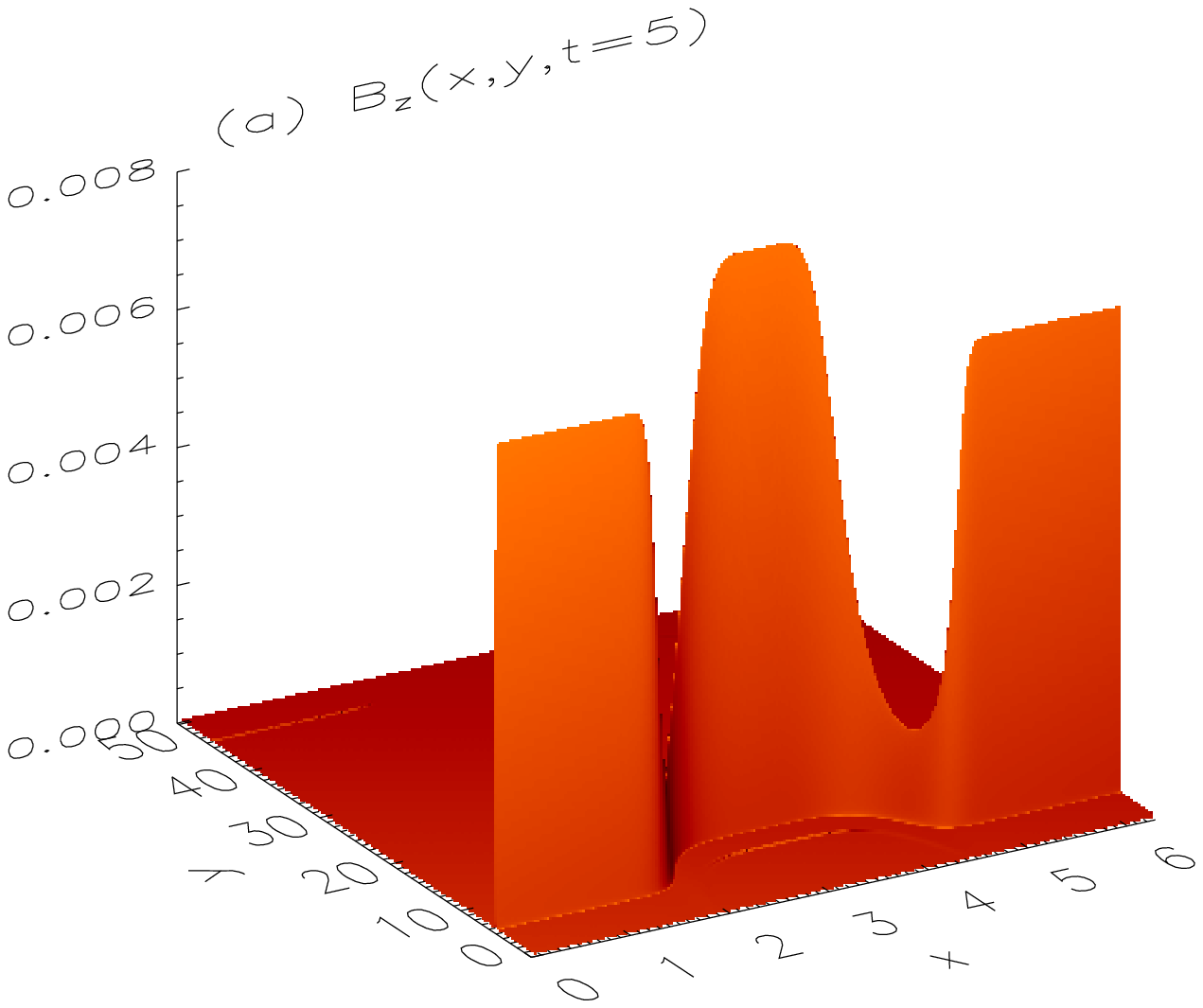,width=8.5cm}
\epsfig{file=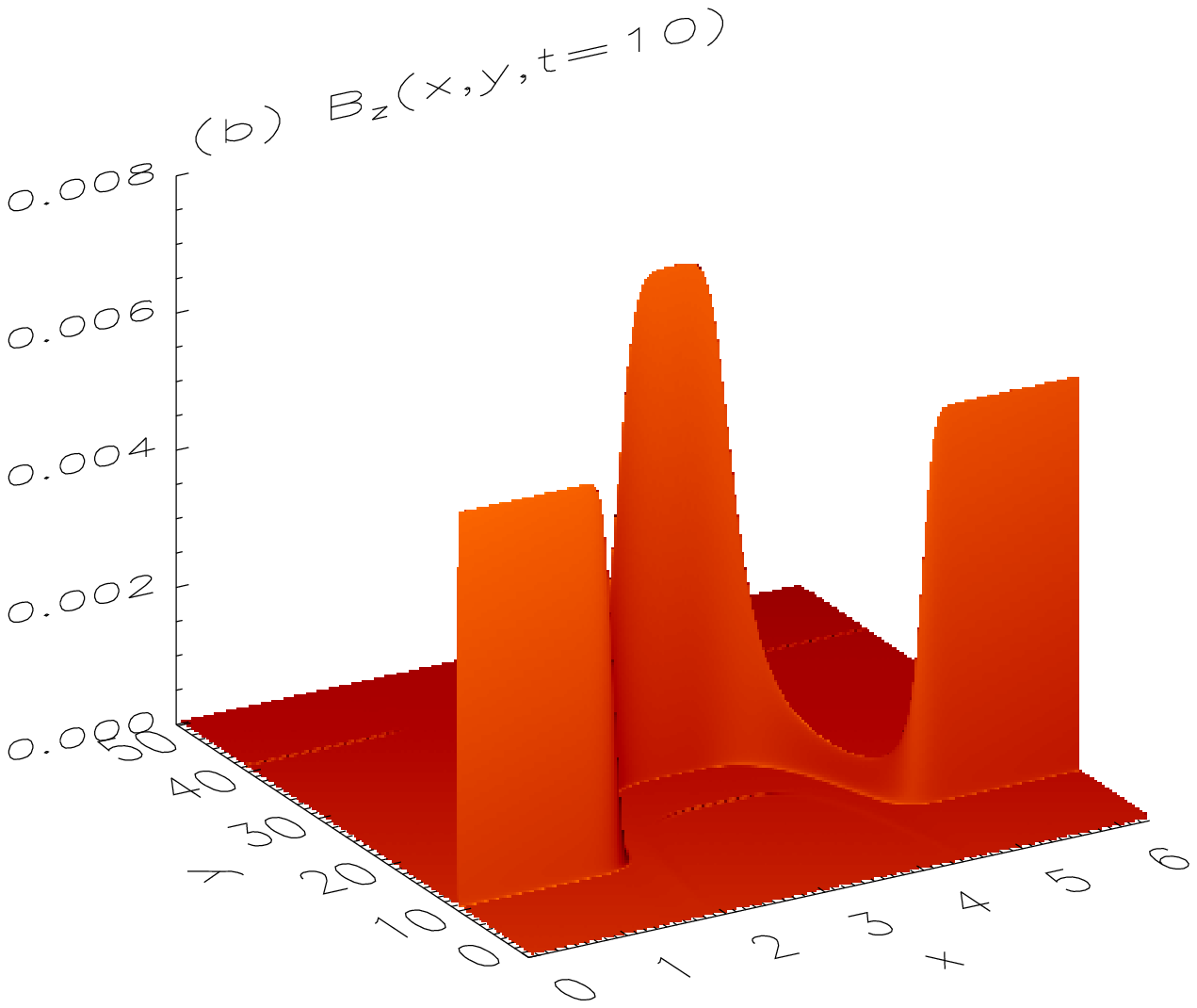,width=8.5cm}
\epsfig{file=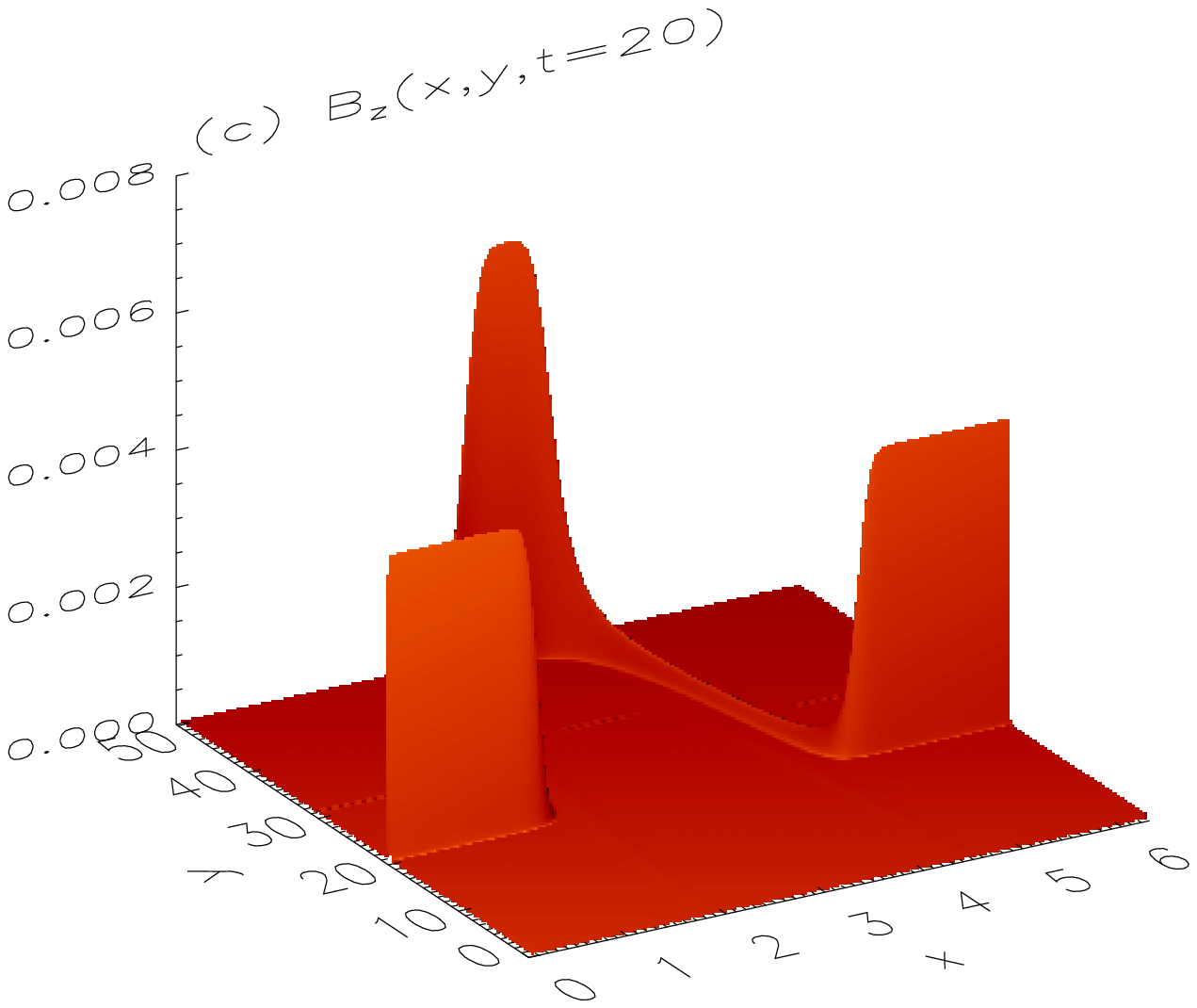,width=8.5cm}
\epsfig{file=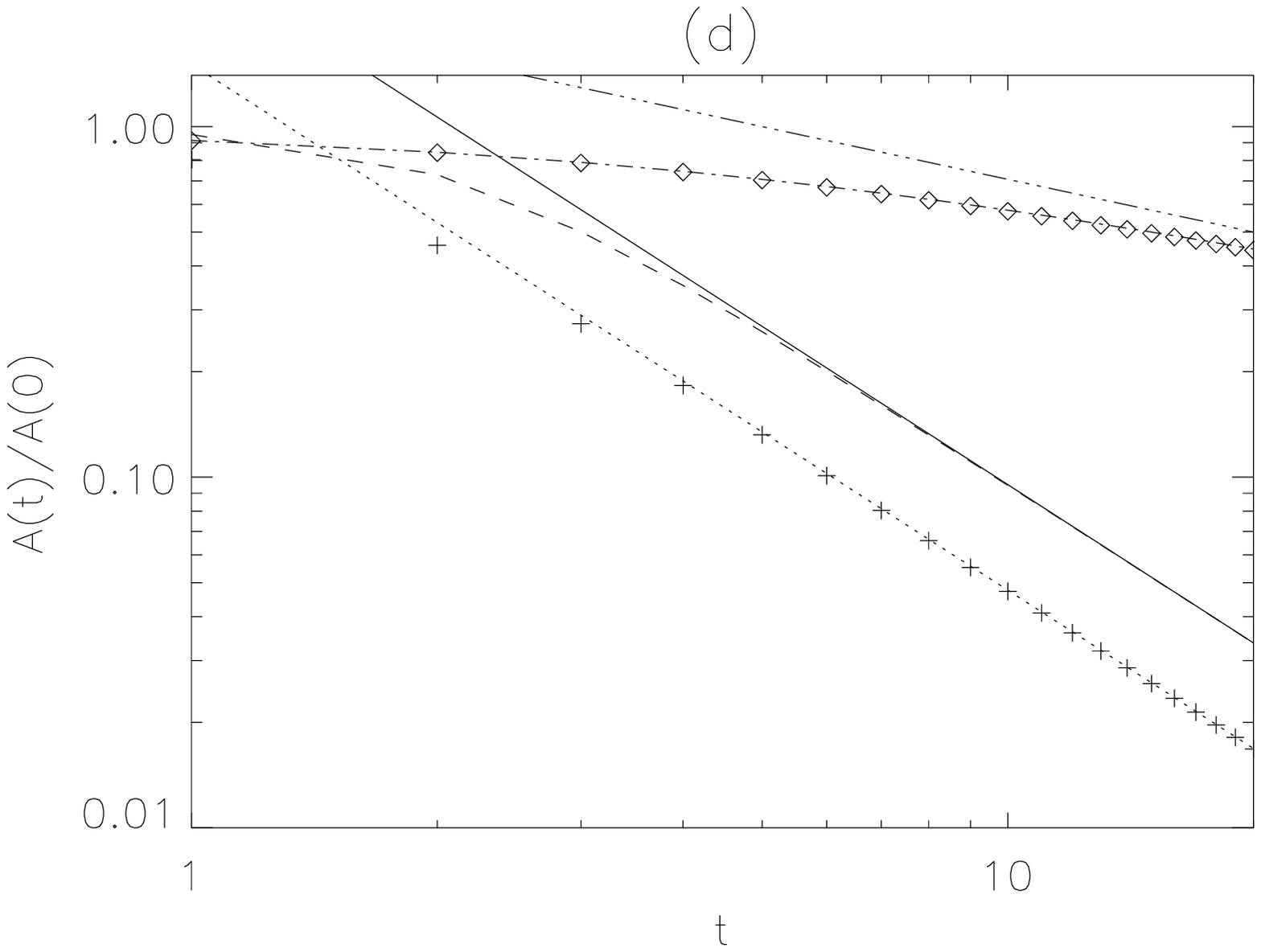,width=8.5cm}
\caption{The same as in Figure \ref{fig3} but for the case of $D=3$.
The dotted line corresponds to the asymptotic solution for large times, Eq.(\ref{22}),
at the
strongest density gradient point $x=(907/3000)\times (2\pi)=1.8996$,
while solid line and dashed curve are kept the same as in Figure \ref{fig3} for comparison. }
\label{fig6}
\end{figure*}

\begin{figure*}
\centering
\epsfig{file=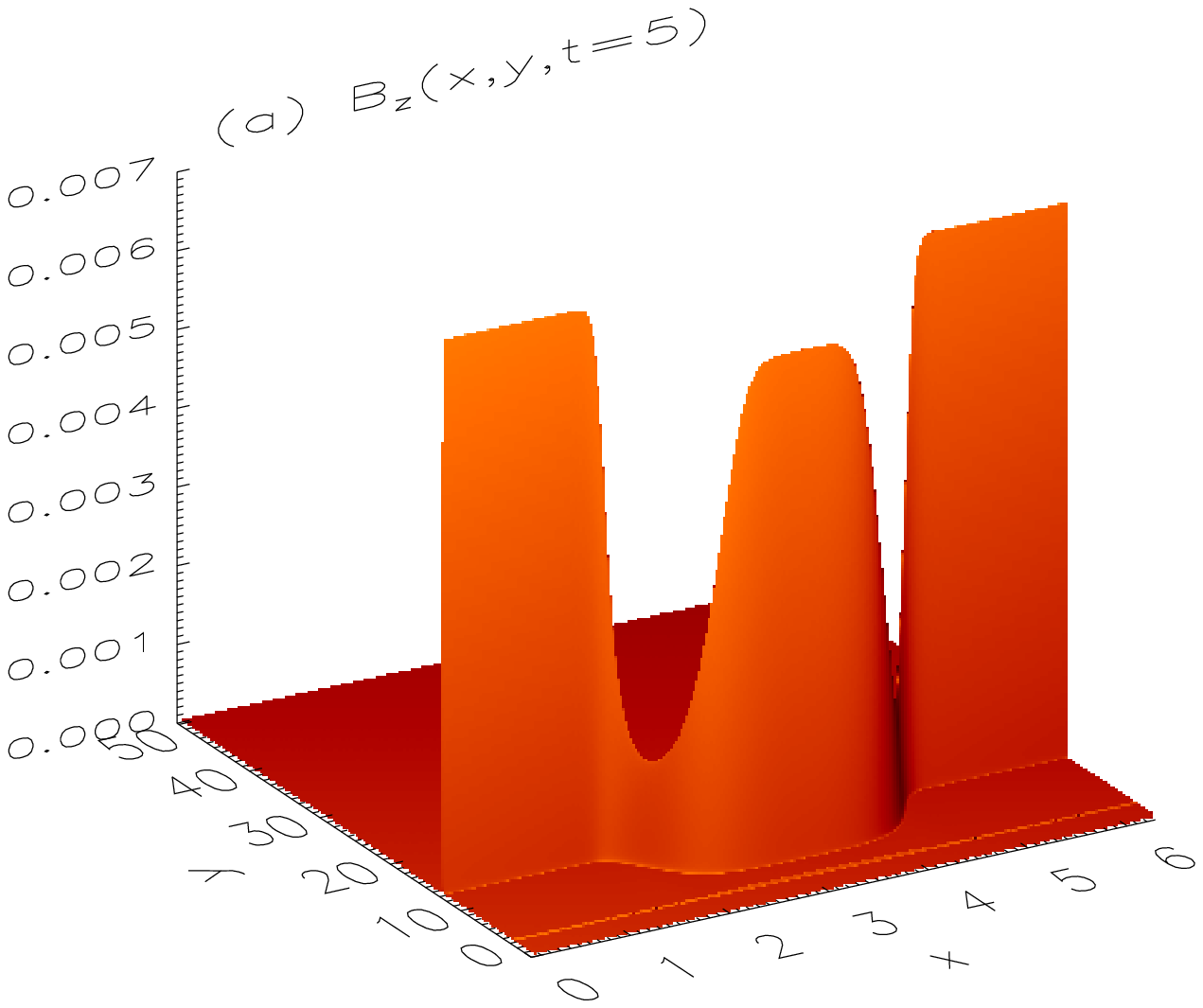,width=8.5cm}
\epsfig{file=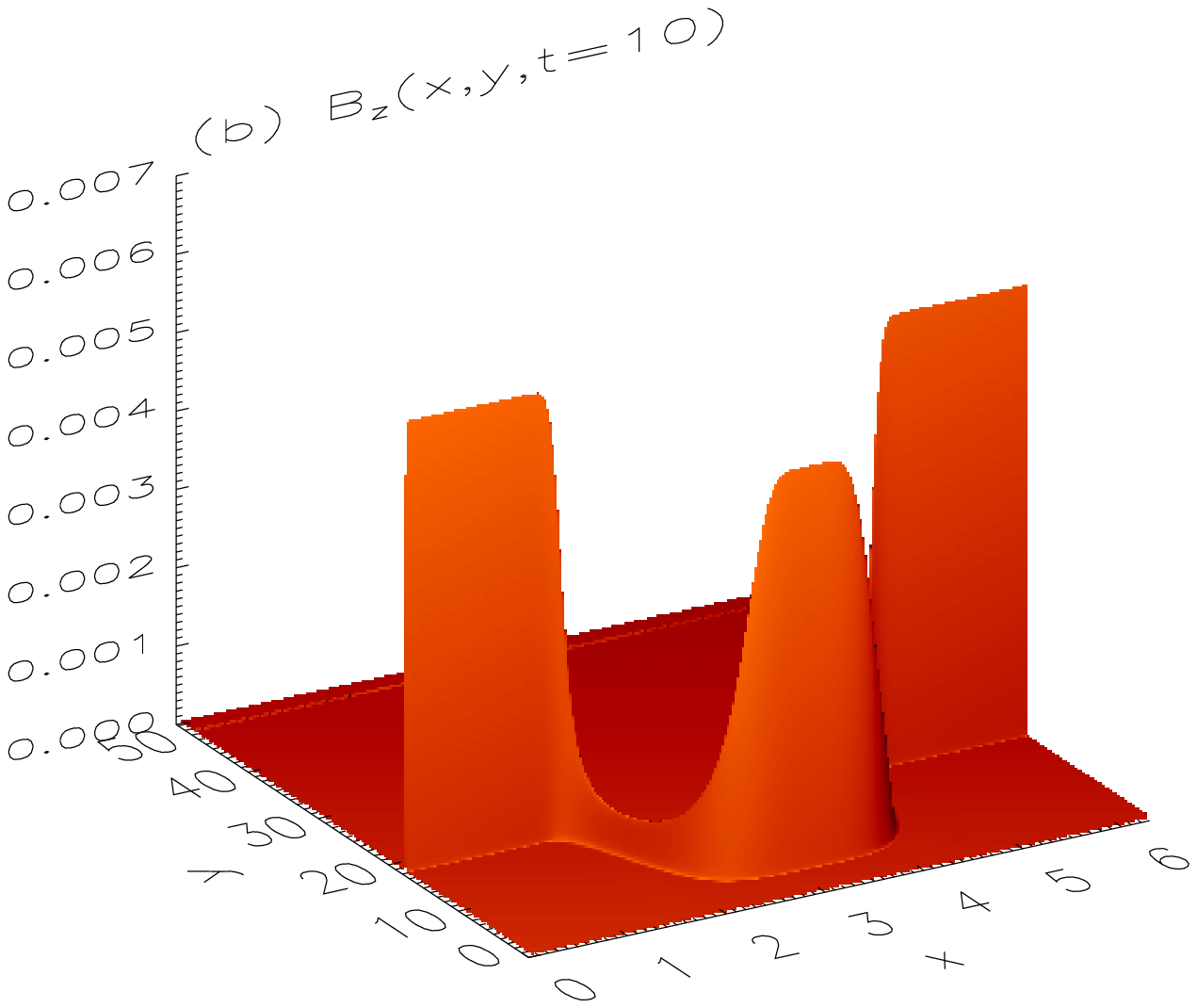,width=8.5cm}
\epsfig{file=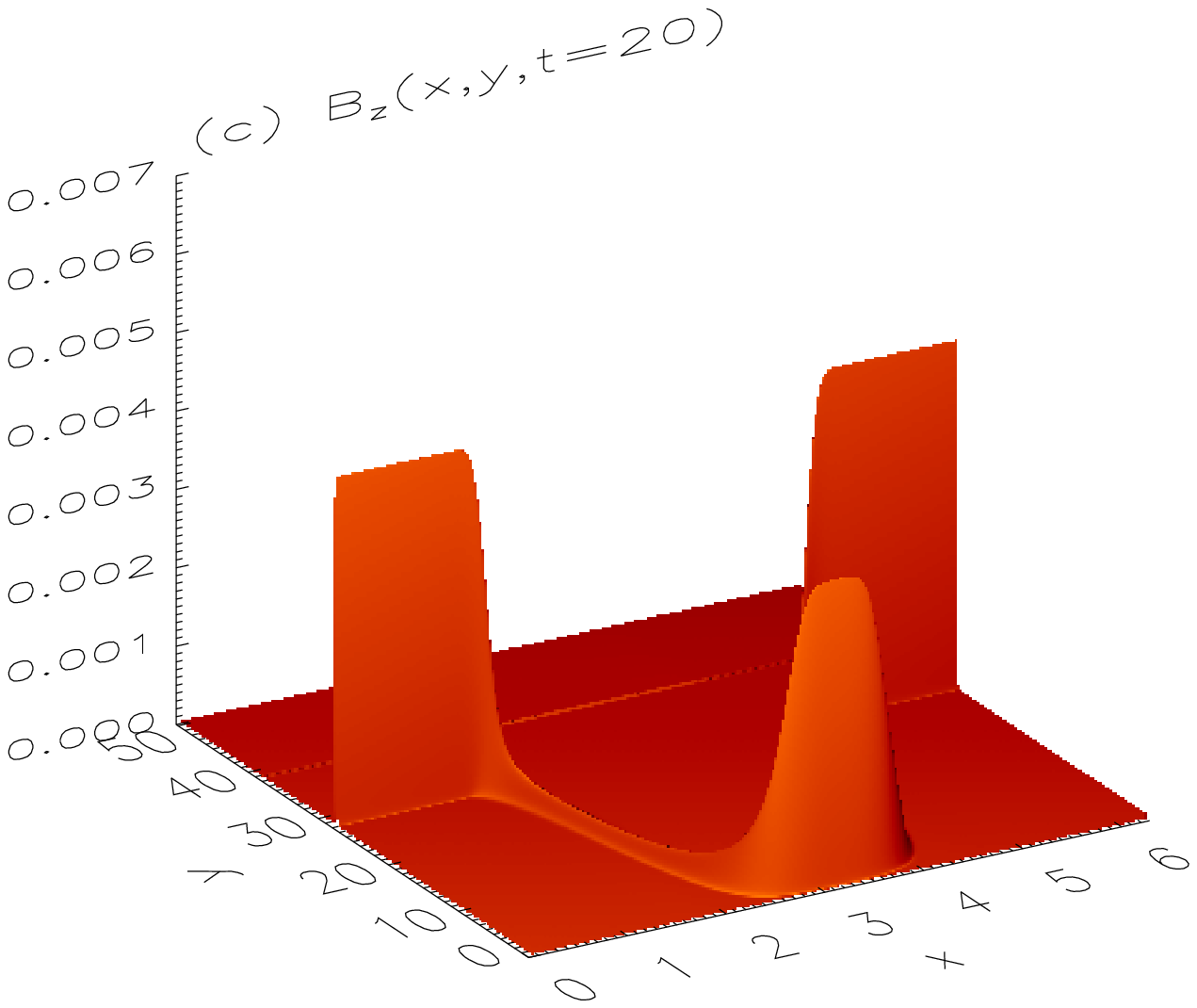,width=8.5cm}
\epsfig{file=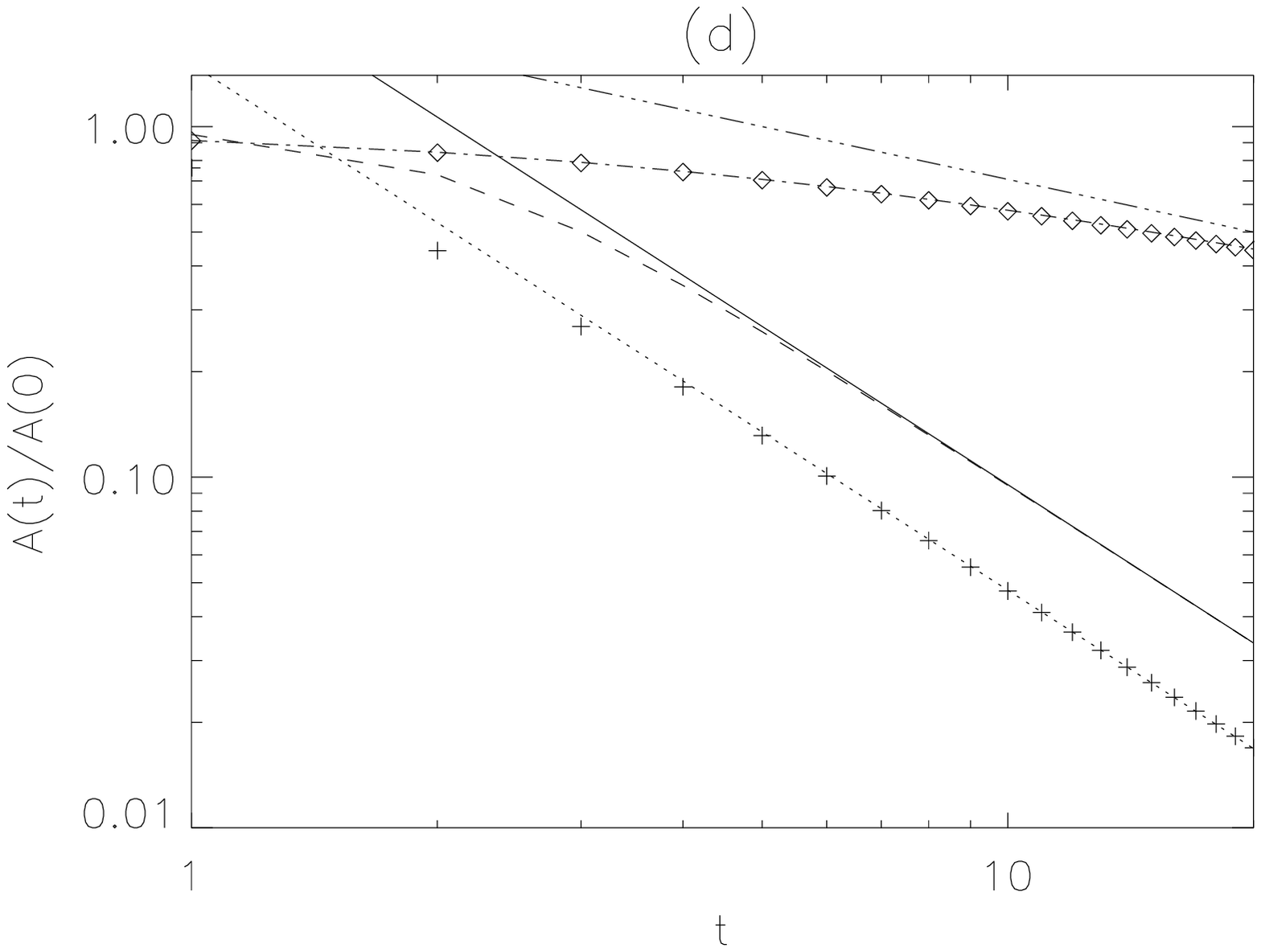,width=8.5cm}
\caption{The same as in Figure \ref{fig6} but for the case of $D=-1$.}
\label{fig7}
\end{figure*}

\section{The model and analytical calculations}

We describe AW dynamics using MHD equations for cold (the background pressure  
is equal to zero, $p_0=0$), incompressible (density perturbation 
is equal to zero, $\rho^\prime=0$) plasma with non-zero resistivity and zero
viscosity ($\eta \not = 0$, $\nu=0$):

\begin{equation}
\frac{\partial \bm V}{\partial t} +(\bm V \cdot \nabla) \bm V=
\frac{(\nabla \times \bm B) \times \bm B}{\mu_0 \rho_0},
\label{1}
\end{equation}
\begin{equation}
\frac{\partial \bm B}{\partial t}= \nabla \times (\bm V \times \bm B) 
-\eta  \nabla \times (\nabla \times \bm B). 
\label{2}
\end{equation}
The background plasma quantities are denoted with subscript $0$, while
perturbation (AW) are denoted with primes as follows:
$\bm V_0=(0,V_0,0)$, $\bm B_0=(0,B_0,0)$, $\rho_0=\rho_0(x)$, and
$\bm V^\prime=(0,0,V_z^\prime)$, $\bm B^\prime=(0,0,B_z^\prime)$. 
The $z$-coordinate is assumed to be an ignorable direction, i.e.
$\partial / \partial z =0$.

Linearly polarised AW is described by the $z$-component of 
Eqs.\ref{1} and \ref{2}:
\begin{equation}
\frac{\partial V_z^\prime}{\partial t} + V_0\frac{\partial V_z^\prime}{\partial y} =
\frac{((\nabla \times \bm B ^\prime) \times \bm B_0)_z}{\mu_0 \rho_0},
\label{3}
\end{equation}
\begin{equation}
\frac{\partial B_z^\prime}{\partial t}= (\nabla \times (\bm V^\prime \times \bm B_0))_z
+ (\nabla \times (\bm V_0 \times \bm B^\prime))_z
+\eta  \nabla^2 B_z^\prime. 
\label{4}
\end{equation}
In Eq.(\ref{4}), the vector identity 
$\nabla \times (\nabla \times \bm B)= \nabla 
(\nabla \cdot \bm B)-\nabla^2 \bm B$ has been used
with the divergence of the magnetic field being zero ($\nabla \cdot \bm B=0$).
Therefore, the system of equations that describes AW dynamics (and dissipation)
is as follows
\begin{equation}
\frac{\partial V_z^\prime}{\partial t} + V_0\frac{\partial V_z^\prime}{\partial y} =
\frac{B_0}{\mu_0 \rho_0}\frac{\partial B_z^\prime}{\partial y},
\label{5}
\end{equation}
\begin{equation}
\frac{\partial B_z^\prime}{\partial t}= B_0\frac{\partial V_z^\prime}{\partial y} 
-V_0\frac{\partial B_z^\prime}{\partial y} 
+\eta  \nabla^2 B_z^\prime. 
\label{6}
\end{equation}
Next, we transform the equations into the frame co-moving with AW plus background flow speeds
with the following coordinates $(x,y,t)\to(\bar x, \xi, \tau)$:
$\bar x =x$, $\xi = y-C_A(x)t-V_0(x)t$ and $\tau = \varepsilon t$ 
(with $\varepsilon \ll 1$).
The derivatives in the new coordinate system are as follows:
\begin{equation}
\frac{\partial }{\partial x} = \frac{\partial }{\partial \bar x} -
(C_A^\prime(x)+V_0^\prime(x))t \frac{\partial }{\partial \xi},
\label{7}
\end{equation}
\begin{equation}
\frac{\partial }{\partial y} =\frac{\partial }{\partial \xi},
\label{8}
\end{equation}
\begin{equation}
\frac{\partial }{\partial t} =  - (C_A(x)+V_0(x)) \frac{\partial }{\partial \xi} 
+\varepsilon \frac{\partial }{\partial \tau}. 
\label{9}
\end{equation}
Applying the transform to the linearized first order system of 
equations (\ref{5}) and (\ref{6}), and algebraically cancelling the terms
that contain $V_0$ yields\begin{equation}
\left(- C_A(x) \frac{\partial }{\partial \xi} 
+\varepsilon \frac{\partial }{\partial \tau} \right)V_z^\prime =
\frac{B_0}{\mu_0 \rho_0}\frac{\partial B_z^\prime}{\partial \xi},
\label{10}
\end{equation}
\begin{eqnarray}
\nonumber
\left(- C_A(x) \frac{\partial }{\partial \xi} 
+\varepsilon \frac{\partial }{\partial \tau} \right)B_z^\prime =
B_0\frac{\partial V_z^\prime}{\partial \xi} + \\
\eta \left(\frac{\partial }{\partial \bar x} -
(C_A^\prime(x)+V_0^\prime(x))t 
\frac{\partial }{\partial \xi}\right)^2B_z^\prime.
\label{11}
\end{eqnarray}
In Eq.\ref{11} in the 
$\nabla^2=\partial^2/\partial x^2 +\partial^2/\partial y^2$
we only kept $\partial^2/\partial x^2$ because
of the phase-mixing $\partial^2/\partial x^2 \gg \partial^2/\partial y^2$.
The next step is to apply operator $-C_A(x)\partial / \partial \xi+
\varepsilon \partial / \partial \tau$ to Eq.(\ref{11}):
\begin{eqnarray}
\left(- C_A(x) \frac{\partial }{\partial \xi} 
+\varepsilon \frac{\partial }{\partial \tau} \right)^2B_z^\prime  = 
B_0\left(- C_A(x) \frac{\partial }{\partial \xi} 
+\varepsilon \frac{\partial }{\partial \tau} \right) 
\frac{\partial V_z^\prime}{\partial \xi} \nonumber \\
 +\eta \left(- C_A(x) \frac{\partial }{\partial \xi} 
+\varepsilon \frac{\partial }{\partial \tau} \right)
\left(\frac{\partial }{\partial \bar x} -
(C_A^\prime(x)+V_0^\prime(x))t 
\frac{\partial }{\partial \xi}\right)^2B_z^\prime. 
\label{12}
\end{eqnarray}
We then apply operator $\partial / \partial \xi$ to Eq.(10)
\begin{equation}
\left(- C_A(x) \frac{\partial }{\partial \xi} 
+\varepsilon \frac{\partial }{\partial \tau} \right) 
\frac{\partial V_z^\prime}{\partial \xi} =
\frac{B_0}{\mu_0 \rho_0}\frac{\partial^2 B_z^\prime}{\partial \xi^2},
\label{13}
\end{equation}
and substitute the latter into Eq.(\ref{12}).
\begin{eqnarray}
\left(- C_A(x) \frac{\partial }{\partial \xi} 
+\varepsilon \frac{\partial }{\partial \tau} \right)^2B_z^\prime =
\frac{B_0^2}{\mu_0 \rho_0}\frac{\partial^2 B_z^\prime}{\partial \xi^2} \nonumber \\
+\eta \left(- C_A(x) \frac{\partial }{\partial \xi} 
+\varepsilon \frac{\partial }{\partial \tau} \right)
\left(\frac{\partial }{\partial \bar x} -
(C_A^\prime(x)+V_0^\prime(x))t 
\frac{\partial }{\partial \xi}\right)^2B_z^\prime. 
\label{14}
\end{eqnarray}
Eq.(\ref{14}) is an equation for $B_z^\prime$
and can be solved analytically using simplifying assumptions
in the asymptotic limit of large time $t / \tau_A \gg 1$.
Here $\tau_A$ is the Alfv\'en time $\tau_A = L / C_A(x)$,
with $L$ and $C_A(x)=B_0/(\mu_0 \rho_0(x))^{0.5}$ being a typical lengthscale 
of the system and Alfv\'en speed, respectively.
Ignoring $\varepsilon^2\ll 1$ order terms, whilst retaining only
$t^2 \gg 1$ order terms in the term proportional to $\eta$, yields
\begin{equation}
- 2 C_A(x)\varepsilon \frac{\partial^2 B_z^\prime }
{\partial \xi \partial \tau} =
-\eta C_A(x) \frac{\partial }{\partial \xi}
(C_A^\prime(x)+V_0^\prime(x))^2 t^2 \frac{\partial^2 B_z^\prime }
{\partial \xi^2}.
\label{15}
\end{equation}
In the above equation, the 
term $C_A(x)^2\partial^2 B_z^\prime/ \partial \xi^2$
algebraically cancels out.

As in \citet{2003A&A...400.1051T}, we now introduce
an auxiliary quantity that has a physical meaning
of slow diffusion time for an AW:
\begin{equation}
s=\frac{\eta (C_A^\prime(x)+V_0^\prime(x))^2 \tau^3}{6 \varepsilon^3}=
\frac{\eta (C_A^\prime(x)+V_0^\prime(x))^2 t^3}{6},
\label{16}
\end{equation}
and the derivative
\begin{equation}
\frac{\partial}{\partial s}=
\frac{2 \varepsilon^3 }{\eta (C_A^\prime(x)+
V_0^\prime(x))^2 \tau^2 }\frac{\partial}{\partial \tau}.
\label{17}
\end{equation}
Using the new notation and after integration by $\xi$, 
Eq.(\ref{15}) reduces to the diffusion
equation
\begin{equation}
\frac{\partial B_z^\prime}{\partial s}=
\frac{\partial^2 B_z^\prime}{\partial \xi^2}.
\label{18}
\end{equation}

Following similar approach as in 
\citet{2003A&A...400.1051T}, Eq.(\ref{18}) can be 
integrated as
\begin{equation}
B_z={\frac {1} {2 \sqrt{\pi s}}} \int_{-\infty}^{+\infty}
\exp \left[{-{\frac{(\xi-\xi')^2}{4 s}}}\right] 
\, B_z(\xi',t=0) \, d \, \xi'.
\label{19}
\end{equation}

Let us substitute a harmonic wave initial condition
$B_z(\xi',t=0)=\exp(i k \xi')$ into Eq.(\ref{19}). The
integration provides a solution
\begin{equation}
B_z= e^{- \eta (C_A^\prime(x)+V_0^\prime(x))^2 t^3 k^2/ 6} \, e^{-i k(y-C_A(x) t-V_0(x) t)}.
\label{20}
\end{equation}
Eq.(\ref{20})
generalises the well known Heyvaerts \& Priest's solution
for the case of shear flow, which is modified by the
following substitution $C_A^\prime(x) \to C_A^\prime(x)+V_0^\prime(x)$.

For a Gaussian pulse of the following mathematical form,
$B_z(\xi',t=0)= \alpha_0 e^{-{\xi'}^2/2 \sigma^2}$, 
its substitution  into Eq.(\ref{19}) 
gives a solution
\begin{eqnarray}
\nonumber
B_z={\frac{\alpha_0}{ \sqrt{1+ 
\eta (C_A^\prime(x)+V_0^\prime(x))^2 t^3/ 3 \sigma^2}}} \\
\times \exp{\left[{-{\frac{[y-(C_A(x)+V_0(x)) t]^2}{2 (\sigma^2+
\eta (C_A^\prime(x)+V_0^\prime(x))^2 t^3/ 3)}}}\right]}, 
\label{21}
\end{eqnarray}
which generalises the solutions that have been obtained 
before \citep{2002RSPSA.458.2307W,2003A&A...400.1051T}.
Here, $\alpha_0=1/5 \sqrt{2 \pi} \sigma$.
In the asymptotic limit of large times, $t$, Eq.(\ref{21}) implies that the
amplitude of AW Gaussian pulse damps as 
\begin{equation}
B_z= \frac{1}{5} \left[2 \pi \eta (C_A^\prime(x)+
V_0^\prime(x))^2/3\right]^{-1/2}\,t^{-3/2}.
\label{22}
\end{equation}

\section{Two-dimensional MHD simulations}

The 2D MHD numerical simulations of this 
work employ Lare2d \cite{2001JCoPh.171..151A} -- 
a Lagrangian remap code that solves non-linear MHD equations. 
Lare2d is second-order accurate 
in space and time. 
The code is available for download from 
\url{http://ccpforge.cse.rl.ac.uk/gf/project/lare2d/}.
Lare2d uses shock viscosity and gradient limiters to  
 capture shock. However, the amplitudes considered 
in this work are weakly non-linear. 
In all our numerical simulations we use
 a 2D box with $3000 \times 24000$ uniform grids in $x$ and $y$
direction, which have alength of $2\pi$  and $16\pi$ in each  
spatial direction, respectively. 
The distance, magnetic field, and density are 
normalised to their background values $L,B_0,\rho_0$.
The velocity and time are 
normalised to the corresponding Alfv\'en values 
$C_A=B_0/\sqrt{\mu_0 \rho_0}$ and $\tau_A=L/C_A$ at $x=0$.
Boundary conditions are periodic in both spatial directions.
In all numerical runs, a normalised, 
uniform magnetic field, of strength unity, is in $y$-direction.
The density has a profile in $x$-direction 
$\rho(x)=1+9\exp(-(x-\pi)^4 )$. 
Therefore, a normalised Alfv\'en speed profile is
\begin{equation}
C_A(x)=1 /\sqrt{\rho_0(x)}=1/\sqrt{1+9\exp(-(x-\pi)^4 )}.
\label{23}
\end{equation}
Plasma beta and 
gravity are set to zero in all numerical runs. For numerical reasons,
plasma beta is actually set to $10^{-8}$, but effectively it is zero.
In all our simulations with AWs at $t=0$ we impose 
a Gaussian pulse which has two components, 
$B_z= 0.01 \exp(-(y-0.5)^2/(2 \times 0.05^2))$,
$V_z= -0.01 \exp(-(y-0.5)^2/(2 \times 0.05^2))/\sqrt{\rho_0(x)}$,
making it a linearly polarised AW packet with  an amplitude of $0.01$. The pulse starts at $y=0.5$ 
and has a width of $\sigma=0.05$. 
Only in Figure \ref{fig7} the pulse starts at 
$y=8$, so that the backflowing middle part stays within the simulation domain.
Plasma viscosity is set to zero, while first 
and second shock viscosity coefficients are set at 0.01 and 0.05
(see \cite{2001JCoPh.171..151A} for further details).
The plasma resistivity is always set to $\hat \eta=5\times 10^{-4}$.
The resistivity is quoted in units of $\mu_0 L C_A$. Therefore, 
$1/ {\hat \eta} =S$ is the Lundquist number.
Plasma flow runs along $y$-direction and 
its mathematical form is given by
\begin{equation}
V_0(x)=D-D /\sqrt{\rho_0(x)}=D-D C_A(x),
\label{24}
\end{equation}
with constant $D=0,1,2,3$ or $D=-1$ controlling the cases of
(i) no flow (usual phase-mixing, 
\citep{2002RSPSA.458.2307W,2003A&A...400.1051T}), 
(ii) flat profile across $x$-coordinate (no phase mixing),
(iii) forward flow, exceeding AW speed, 
(iv) stronger forward flow, further exceeding AW speed,
and 
(v)  backward flow, respectively.
Each numerical run takes about five hours on 256 CPU computing cores
on a COSMA-5 supercomputer  \url{http://dirac.ac.uk/Resources.html}.
The different examples are illustrated in Figure \ref{fig1}.
It follows from Figure \ref{fig1}(a) that for the chosen
set of parameters, 
Alfv\'en wave ($C_A(x)$, dashed curve)  in the over-dense region $1.5 < x< 4.5$ lags
 with $\min(C_A(x))=0.3162$. For $D=1$,
the background plasma flow ($V_0(x)$, dotted curve) speed has a maximum 
$\max(V_0(x))=1-0.3162=0.6838$ in the same region,
such that 
the sum of the two $C_A(x)+V_0(x)=1$
(solid curve) for all $x$.
The same follows from the analytical expressions
Eq.(\ref{23}) and (\ref{24}), i.e. for $D=1$ the $C_A(x)$ terms
cancel out.
Thus, $D=1$ case
corresponds to a flat profile across $x$-coordinate (no phase-mixing).
In other words the forward flow completely counteracts
the wave-front stretching because of the variation of Alfv\'en speed
with the transverse coordinate.
Figure \ref{fig1}(b) is for $D=2$, forward flow, whose speed  
exceeds AW speed reduction,
such that $\max(V_0(x))=1.3675$  and 
$\max(C_A(x)+V_0(x))=1.6838$.
The velocity sum difference between over-dense and peripheral 
regions in Figure \ref{fig1}(b) is $1.6838-1=0.6838$, 
the same as the Alfv\'en speed decrease $1-0.3162=0.6838$.
In other words, in Figure \ref{fig1}(b), the solid and
dashed lines are perfectly symmetrical with respect to $y=1$.
Figure \ref{fig1}(c) is for $D=3$, even stronger forward flow, 
such that 
$\max(V_0(x))=2.0513$  and 
$\max(C_A(x)+V_0(x))=2.3675$.
Figure \ref{fig1}(d) is for $D=-1$ backward flow, such that
$\min(V_0(x))=-0.6838$  and 
$\min(C_A(x)+V_0(x))=-0.3675$.
The velocity sum difference between over-dense and peripheral 
regions in Figure \ref{fig1}(c) is $2.3675-1=1.3675$,
the same as that in Figure \ref{fig1}(d)  $1-(-0.3675)=1.3675$. 
In other words, in Figure \ref{fig1}(c)  and 
Figure \ref{fig1}(d), the solid curves have
the same gradient strength, which is also clear visually.

In Figure \ref{fig2} we plot 
the difference between background flow speed at time $t$ as a function of
$x$-coordinate,
$V_0(x,y=y_{max}/2,t)$ and its initial value at $t=0$, $V_0(x,y=y_{max}/2,0)$,
i.e. $\Delta V_y \equiv V_0(x,y=y_{max}/2,t)-V_0(x,y=y_{max}/2,0)$ for different times. 
 The dashed curve is for $t=5$, while the 
 dotted curve is for $t=10$ and solid  for $t=20$. 
In this numerical run the  background flow 
is the fastest of all the
performed numerical runs with $D=3$ (as in panel (c) from 
Figure \ref{fig1}). We gather from Figure \ref{fig2} that, 
 by the end simulation time of $t=20$, the flow speed difference is 
quite small, $\approx 0.005$. It is even smaller for earlier 
times and/or smaller values of $D$. 
Therefore in Figure \ref{fig2}, we demonstrate that the 
background flow in the absence of
the AW pulse does not break up.

Figure \ref{fig3} shows the numerical run results for the case of
no background flow with $D=0$. This corresponds to a similar set-up
studied in \citet{2002RSPSA.458.2307W,2003A&A...400.1051T}
or more recently in \citet{2014PhPl...21e2902T}, as the
numerical code benchmarking exercise.
Figure \ref{fig3} shows shaded surface plots of the
AW, i.e.
$B_z(x,y)$ at different times. 
Panel (a) is for $t=5$, (b) for $t=10$, and (c) for $t=20$.
Panel (d)  shows the
time evolution of AW amplitude, normalised to its initial value. 
The  solid line corresponds to the asymptotic solution 
for large times, Eq.(\ref{22}). 
A more general analytical form 
Eq.(\ref{21}) is plotted with a dashed curve. 
Crosses and open diamonds represent numerical simulation results in the
strongest density gradient point $x=(907/3000)\times (2\pi)=1.8996$ and 
away from the gradient $x=(1/3000)\times (2\pi)=0.0021$ (the first grid 
cell in $x$-direction), respectively.
The Dash-triple-dotted line corresponds 
to the asymptotic solution for large times, Eq.(\ref{31}),
which is independent of $x$. 
A more general analytical form of 
Eq.(\ref{30}) is plotted with a dash-dotted curve. It is also independent of
$x$ because, at the peak of the pulse, the value of the exponent is unity. 
At $t=0$ (snapshot not shown here) AW is initially flat as in, 
e.g. Figure \ref{fig4}(a), but without the hump in the middle and 
instead it is located at $y=0.5$, according to the
initial conditions given above.
The Alfv\'en wavefront quickly damps (the shaded surface disappears from 
\ref{fig3}(a) to \ref{fig3}(c) in the density
inhomogeneity regions $x\approx1.5-2.5$ and $x\approx3.5-4.5$, where 
the wave fronts distort strongly. The derivative 
of the Alfven speed, $C^\prime_A(x)$,
in $x$-direction, which enters Eqs. (\ref{21}) and (\ref{22}), 
with $V_0=0$ and  $V^\prime_0=0$ as $D=0$,
 is responsible for the fast damping of the AW.
Away from the density gradient regions, a much slower
dissipation takes place. The latter  
is hardly noticeable on the timescales concerned ($0<t<20$).
 Away from the density gradient regions, the analytical solutions
Eqs. (\ref{30}) and (\ref{31}) seem to match  the corresponding 
numerical solution (open diamonds) well.

In Figure \ref{fig4} we present numerical run 
results in a similar manner to
Figure \ref{fig3}, except that $D=1$. 
 Note that in panel \ref{fig4}(d) we 
 have adjusted the plot $y$-range to $0.4-1.4$ to see the amplitude
behaviour more clearly. 
At all times, we see that the AW front remains flat
across $x$-coordinate, and the phase-mixing effect is absent. Thus, 
for over-dense plasma structures, which have smaller $C_A$ 
compared to the surrounding plasma, the plasma flow that is confined
to this structure and running in the same direction as the AW, reduces the effect of 
phase-mixing because, on the edges of the structure, 
$C_A^\prime$ and $V_0^\prime$ have the opposite signs. 
In fact, we deduce from Eqs.(\ref{23}) and (\ref{24})
that for $D=1$, $C_A(x)+V_0(x)=1=$const and therefore
$C_A^\prime(x)+V_0^\prime(x)=0$. Thus, the rather slow
wave damping is not due to phase-mixing but to the usual
Spitzer resistivity (note the crosses nearly 
coincide with open diamonds in 
Figure \ref{fig4}(d)).
 This  small mismatch, of the order of $0.0008$, 
could be due to the two following  reasons:
(i) AW has two components ($V_z$,$B_z$) and, as stated above,
the initial condition for $V_z$ is
$V_z= -0.01 \exp(-(y-0.5)^2/(2 \times 0.05^2))/\sqrt{\rho_0(x)}$.
Thus, despite the fact that the wave front is flat, because of the
presence of the flow, the density is still a function of transverse
coordinate $x$;
(ii) for the latter reason, the non-linearity will damp the wave front in a slightly
different way. 
Although, our analytical problem is linear, we  validate it with a fully non-linear
MHD code Lare2d. 
In Figure \ref{fig2}, the maximal flow difference $\Delta V_y$ is approximately $0.005$.
Thus, the small mismatch of $\approx 0.0008$ could be attributed mostly
to the non-linearity of the numerical code.
We stress that 
the plasma resistivity is always set to $\hat \eta=5\times 10^{-4}$ everywhere.
Consequently, the small difference is not due to the weak
(logaritm of plasma parameter) density dependence of the Spitzer resistivity being
affected by the background density profile. 
Thus, the wave damps only by Spitzer (uniform) resistivity 
not by phase-mixing.
This is the result of the co-directional 
flow, which reduces the wave
front stretching in the transverse direction to zero, while the front remains
flat across $x$ at all times. 

Figure \ref{fig5} shows numerical run results  similar to those found in 
Figure \ref{fig3}, except for $D=2$.
The choice of value of $D$ is such that  wave front in Figure 
\ref{fig5}(a)--\ref{fig5}(c) now bends forward instead of backward
compared to Figure \ref{fig3}(a)--\ref{fig3}(c).
We notice that Figure \ref{fig5}(d) seems identical to 
\ref{fig3}(d). This is because the AW-amplitude damping, using the phase-mixing
formula Eqs.(\ref{22}), contains the square of the sum of
the Alfv\'en and flow speed derivatives.
In the case of Figure \ref{fig3}, the wave front derivative
is negative in the density
inhomogeneity regions $x\approx1.5-2.5$,  
while in the case of Figure \ref{fig5}, it is positive.
However, since the derivative is squared, the net effect is the same --
the case without a flow and with forward flow with $D=2$, the AW damping
is the same.

In Figure \ref{fig6} we present numerical 
run results similar to those found  in 
Figure \ref{fig3}, except for $D=3$. This is 
the strongest flow case we consider
and it  demonstrates that fast flow can induce
wave front stretching to the extent that it exceeds
the usual effect of phase-mixing without a flow.
The latter can be clearly seen by looking at the crosses and the dotted line in 
Figure \ref{fig6}(d), which appear lower than the solid line, thus
indicating a stronger  damping, as prescribed by
Eqs.(\ref{21}) and (\ref{22}).

Figure \ref{fig7} depicts the numerical run results  as in 
Figure \ref{fig6}, except for $D=-1$. This corresponds to the back-flow case, i.e.
the AW propagates in the opposite direction to the background flow.
Compared to the case
without the flow, Figure \ref{fig3}(a)--\ref{fig3}(c), in Figure \ref{fig7}(a)--\ref{fig7}(c) we see the
AW stretching is stronger and therefore the wave damping via  the phase-mixing is
faster. However, the damping is the same as in the case where $D=3$.
Therefore Figure \ref{fig7}(d) appears identical to \ref{fig6}(d).

To summarise, based on Eqs.(\ref{23}) and (\ref{24}),
$C_A(x)+V_0(x)= D+(1-D)C_A(x)$, and therefore,
$(C_A^\prime(x)+V_0^\prime(x))^2=(1-D)^2C_A^\prime(x)^2$.
The latter 
prescribes the AW damping via
phase-mixing in Eqs.(\ref{21}) and (\ref{22}).
Thus, $(C_A^\prime(x)+V_0^\prime(x))^2=$
\begin{itemize}
\item[] $C_A^\prime(x)^2$ for  $D=0$, (Figure \ref{fig3}),
\item[] $0$ for $D=1$, (Figure \ref{fig4}),
\item[] $C_A^\prime(x)^2$ for  $D=2$, (Figure \ref{fig5}),
\item[] $4C_A^\prime(x)^2$ for $D=3$, (Figure \ref{fig6}),
\item[] $4C_A^\prime(x)^2$ for  $D=-1$, (Figure \ref{fig7}).
\end{itemize}
Thus, the AW damping via phase-mixing is the same in the 
cases of $D=0$ and $D=2$, and similarly
for cases $D=3$ and $D=-1$. For  $D=1$, the phase-mixing effect is zero. 
Note that 
phase-mixing always acts in addition to the usual 
(homogeneous plasma case) resistive damping,
which diffuses the AW pulse in the other spatial direction (in $y$).
This can be seen by the slight broadening of the pulse in $y$ direction 
in Figures \ref{fig3}--\ref{fig7}.

To quantify the AW Gaussian pulse damping in the homogeneous
plasma case, we now consider a one-dimensional (1D) analogue of our
2D model by suppressing the variation in $x$-direction.
Therefore we consider a 1D AW pulse moving along $y$ in homogeneous plasma.
In this case Equation (\ref{14})
is replaced by
\begin{eqnarray}
\left(- C_A(x) \frac{\partial }{\partial \xi} 
+\varepsilon \frac{\partial }{\partial \tau} \right)^2B_z^\prime =
\frac{B_0^2}{\mu_0 \rho_0}\frac{\partial^2 B_z^\prime}{\partial \xi^2} \nonumber \\
+\eta \left(- C_A(x) \frac{\partial }{\partial \xi} 
+\varepsilon \frac{\partial }{\partial \tau} \right)
\frac{\partial^2B_z^\prime }{\partial \xi^2}. 
\label{25}
\end{eqnarray}
Here, $\left({\partial }/{\partial \bar x} -
(C_A^\prime(x)+V_0^\prime(x))t 
{\partial }/{\partial \xi}\right)^2$ 
operator has been replaced by 
${\partial^2 }/{\partial y^2}={\partial^2 }/{\partial \xi^2}$
because in the 2D phase-mixing case in Eq.(\ref{14}),
we set $\partial^2/\partial x^2 \gg \partial^2/\partial y^2$,
while in 1D case $\partial/\partial x=0$. This means that, out of the
$\nabla^2$ operator, we retain ${\partial^2 }/{\partial y^2}$.
Following a similar procedure, as described above,
the equivalent form of Eq.(\ref{15}) is now
\begin{equation}
- 2 C_A\varepsilon \frac{\partial^2 B_z^\prime }
{\partial \xi \partial \tau} =
-\eta C_A \frac{\partial }{\partial \xi}
\frac{\partial^2 B_z^\prime }{\partial \xi^2}.
\label{26}
\end{equation}
We now introduce
an auxiliary quantity,
\begin{equation}
s_1=\frac{\eta \tau}{2 \varepsilon}=\frac{\eta t}{2}.
\label{27}
\end{equation}
Using the new notation and, after integration by $\xi$, 
Eq.(\ref{26}) reduces to the diffusion
equation
\begin{equation}
\frac{\partial B_z^\prime}{\partial s_1}=
\frac{\partial^2 B_z^\prime}{\partial \xi^2}.
\label{28}
\end{equation}
As above, Eq.(\ref{28}) can be 
integrated using Eq.(\ref{19}).
For a Gaussian pulse of the following mathematical form,
$B_z(\xi',t=0)= \alpha_0 e^{-{\xi'}^2/2 \sigma^2}$, 
its substitution  into Eq.(\ref{19}) 
yields
\begin{equation}
B_z^\prime=\frac{\alpha_0}{\sqrt{1+ 
2s_1/ \sigma^2}}
 \exp{\left[{-{\frac{\xi^2}{2 (\sigma^2+
2s_1)}}}\right]}.
\label{29}
\end{equation}
Substituting the definition of $s_1$ provides the desired
solutions
\begin{equation}
B_z=\frac{\alpha_0}{\sqrt{1+\eta t / \sigma^2}}
\exp{\left[{-{\frac{[y-(C_A(x)+V_0(x)) t]^2}{2 (\sigma^2+
\eta t)}}}\right]}, 
\label{30}
\end{equation}
and its
asymptotic limit of large times: 
\begin{equation}
B_z= \frac{1}{5} \left[2 \pi \eta \right]^{-1/2}\,t^{-1/2}.
\label{31}
\end{equation}
In the case of homogeneous plasma resistive
damping of the Gaussian AW pulse, the amplitude damps as $t^{-1/2}$
compared to $t^{-3/2}$ for the 2D phase-mixing.
In the uniform density regions, the analytical solutions
Eqs. (\ref{30}) and (\ref{31}) seem to match  the corresponding 
numerical solution (open diamonds) in all Figures \ref{fig3}-\ref{fig7} well.

\section{Conclusions}

This paper uses analytical calculations, corroborated by MHD
simulations,
to demonstrate that, when a flow is present, 
mathematical expressions for the Alfv\'en wave
damping via phase-mixing are modified by the
following substitution $C_A^\prime(x) \to C_A^\prime(x)+V_0^\prime(x)$.
In uniform magnetic field and over-dense plasma 
structures, 
in which $C_A$ is smaller
compared to the surrounding plasma, the flow, which is confined
to this structure,
and in the same direction as the AW, reduces the effect of 
phase-mixing.
This is because, on the edges of the structure, 
$C_A^\prime$ and $V_0^\prime$ have opposite signs. As a result of this, 
the AW damping, 
via phase-mixing, is  slower when compared to the case without the flow.
For example, in the over-dense plasma structures with
density inside ten times higher than outside,
the co-directional with the wave flow with $\approx 0.7 C_A$ 
in the middle of the over-density
(see Figure \ref{fig1}(a) and \ref{fig4}) can
reduce the phase-mixing effect to zero.
This is the consequence of the co-directional 
flow reducing the wave
front stretching in the transverse direction to zero. 
Conversely, the
counter-directional flow increases the wave
front stretching in the transverse direction, therefore making
the phase-mixing effect more effective
(see Figure \ref{fig1}(d) and \ref{fig7}) compared to the case
without the flow.
The flows with $\approx 1.4 C_A$ and  $\approx 2 C_A$
in the middle of the over-density, make the wave front
 go faster in the over-dense region compared to the surrounding
plasma (see Figures \ref{fig1}(b),\ref{fig1}(c), \ref{fig5} and \ref{fig6}).
In the case without the flow (Figure \ref{fig3}),
the  wave front
is slower in the over-dense region compared to the surrounding
plasma. The 
dissipation of the AW via phase-mixing is:
(i) the same for
flows with $\approx 1.4 C_A$ as for those without the flow (although 
in both cases the wave front is
bent forward and backward, respectively) and  
(ii) larger for  case of $\approx 2 C_A$ (which is probably unrealistic
observationally, but is presented for completeness), as quantified by 
Eq.(\ref{22}).
We stress that the result is generic and is applicable
to different laboratory or astrophysical plasma systems
where flows, density inhomogeneity across the background
magnetic field, and AW resistive dissipation are all present.
Nonetheless, we apply our findings to address the question why over-dense 
solar coronal open 
magnetic field structures 
are cooler than the background plasma.
Since observations show that the over-dense OMFS are cooler 
than the surrounding plasma, and that they
are in regions where Doppler line-broadening  is consistent with bulk plasma
motions, e.g. the AW, we show that,
if over-dense solar coronal OMFS are heated by AW damping via phase-mixing,
 the co-directional with the wave plasma flow in them
reduces the phase-mixing-induced heating, thus providing an explanation
for why they are cooler than the surrounding plasma. 

As mentioned in the introduction, and reiterated here,
  there is currently a disagreement 
 as to whether CPPs are the source of fast solar wind (see related 
discussion and references in \citet{1997SoPh..175..393D}).
Some observations claim that the source of the fast solar wind
is the inter-plume region. \citet{2000ApJ...531L..79G}
present a spectroscopic study of the ultraviolet coronal emission in a polar hole. 
They identify
the inter-plume lanes and background coronal hole regions as the 
channels in which the fast solar wind is preferentially accelerated.
i.e. outside the plume the speed is higher than in the plume.
We stress, however, that these observations only present
three measurements of the flow speed: two on either side of the plume,
where flow speed is found faster and one inside where flow speed is 
slower. This does not preclude a possibility that on the edges of the
plume the flow speed falls to zero and thus $V_0^\prime(x)>0$ inside
the plume. Since, in the over-dense structures with a uniform magnetic field
$C_A^\prime(x)<0$, the effect of the flow counteracting the phase-mixing is still
a viable possibility.

\citet{2015SoPh..tmp....8H} explored the changes in coronal 
non-thermal velocity (i.e. bulk flows or AWs) measurements 
at the poles from solar minimum to solar maximum, using Hinode EUV Imaging 
Spectrometer (EIS) data. They find that, although the intensity in the corona at the 
poles does tend to increase with the cycle, there are no significant 
changes in the $V_{non-thermal}$ values. The locations of enhanced $V_{non-thermal}$ values that they
measure do not always have a counterpart in intensity, and they are 
sometimes located in weak emission regions.
The next logical step in corroborating our theory would be
to check whether there is a correlation between temperature and non-thermal velocity
in over- or under-dense OMFS. 
Care should be taken however in interpreting the observational
results. We considered over-dense structures (e.g. coronal plumes) and
found that, co-directional with the wave flows reduce phase-mixing and hence 
reduce heating/wave-dissipation, such that they  appear cooler than their surroundings. 
Obviously, in under-dense structures, where
AW phase speed inside the structure is higher, hence 
co-directional with the wave flows increase the  phase mixing
and thus increase wave-dissipation/temperature.
This would result in a positive correlation between 
proton temperature and solar wind speed \cite{1994JGR....9921481T}; 
(Horbury \& Matteini, priv. comm., manuscript submitted for publication),
i.e. faster streams are also hotter.
Typically during periods of fast solar wind ($V>700$ km s$^{-1}$), with fast streams
($V=150-200$ km s$^{-1}$), 
both the total magnetic field and the density are  
constant. Thus  Alfv\'en speed is constant across the streams. 
\citet{2015ApJ...802...11M} show 
that in terms of the turbulence, the system appears to be 
in a local equilibrium, where there are no jumps in 
particle/field energies. 
This behaviour suggests that the turbulence has evolved 
to a stage where the system is in equilibrium, making 
fast streams and background homogeneous.
However, assuming that at the origin there were independent 
steams of plasma, with different physical properties, 
it is plausible that  Alfv\'en speed is  
different in different streams.
It is likely that the condition $|B|=const$ \citep{2015ApJ...802...11M}
is the result of the relaxation of the turbulence 
and not the initial configuration, where jets could 
have had different $B_0$, which has subsequently been smoothed out.
The next logical step would be  to
do a careful pressure-balance calculation, taking 
into account the temperature changes.

Thus, based on our
model,
the ultimate factor for interpreting 
the observations is dependent on whether Alfv\'en speed
(which is a combination of both density and the magnetic field)
is smaller or larger than in the
surrounding plasma. In summary, 
when plotting
a graph of coronal 
non-thermal velocity, $V_{non-thermal}$, i.e.  the 
background flow speed, versus temperature inside the
structure, $T$, based on, e.g. EIS (Hinode)/ AIA (Solar Dynamics Observatory) 
observations in the solar corona or
Helios observations in fast solar wind streams, the model predicts:
\begin{itemize}
\item[] a positive correlation of $V_{non-thermal}$ with $T$ in the case of
structures in which Alfv\'en speed is larger compared to the surrounding plasma;
\item[] anti-correlation of $V_{non-thermal}$ with $T$ in the case of
structures in which Alfv\'en speed is smaller compared to the surrounding plasma 
(hence the title of this paper).
\end{itemize}
These conclusions are   based on  natural assumptions
that (i) AW phase-mixing has a major role to play in heating these 
structures and (ii) that the flow is forward (co-directional)
with the AW (i.e. solar wind). There is also a caveat that in the above
correlation, $V_{non-thermal}$ means background flow rather than AW or turbulent motions, 
and that the disentangling of the two maybe difficult.
This is maybe easier on the solar disk rather than on the limb, because measuring
Doppler shifts should enable us to differentiate between  regular (up- or down-) flows from
bulk turbulent motions, which only manifest themselves via the line broadening.

\begin{acknowledgements}

Computational facilities: Astronomy
Unit, Queen Mary University of London and UK's STFC Dirac-2 HPC via UKMHD
consortium. The author was
financially supported by STFC consolidated Grant No.
ST/J001546/1 and Leverhulme Trust Research Project Grant No.
RPG-311. 
The author would also like to thank Dr. T.R. Arter for providing
crucially important IT support.
\end{acknowledgements}

\bibliographystyle{aa}

\begin{thebibliography}{24}
\expandafter\ifx\csname natexlab\endcsname\relax\def\natexlab#1{#1}\fi

\bibitem[{{Arber} {et~al.}(2001){Arber}, {Longbottom}, {Gerrard}, \&
  {Milne}}]{2001JCoPh.171..151A}
{Arber}, T.~D., {Longbottom}, A.~W., {Gerrard}, C.~L., \& {Milne}, A.~M. 2001,
  Journal of Computational Physics, 171, 151

\bibitem[{{Aschwanden}(2005)}]{2005psci.book.....A}
{Aschwanden}, M.~J. 2005, {Physics of the Solar Corona. An Introduction with
  Problems and Solutions (2nd edition)} (Springer-Verlag Berlin Heidelberg,
  chapter 9.4)

\bibitem[{{Cirtain} {et~al.}(2007){Cirtain}, {Golub}, {Lundquist}, {van
  Ballegooijen}, {Savcheva}, {Shimojo}, {DeLuca}, {Tsuneta}, {Sakao}, {Reeves},
  {Weber}, {Kano}, {Narukage}, \& {Shibasaki}}]{2007Sci...318.1580C}
{Cirtain}, J.~W., {Golub}, L., {Lundquist}, L., {et~al.} 2007, Science, 318,
  1580

\bibitem[{{De Moortel} {et~al.}(2000){De Moortel}, {Hood}, \&
  {Arber}}]{2000A&A...354..334D}
{De Moortel}, I., {Hood}, A.~W., \& {Arber}, T.~D. 2000, \aap, 354, 334

\bibitem[{{Deforest} {et~al.}(1997){Deforest}, {Hoeksema}, {Gurman},
  {Thompson}, {Plunkett}, {Howard}, {Harrison}, \&
  {Hassler}}]{1997SoPh..175..393D}
{Deforest}, C.~E., {Hoeksema}, J.~T., {Gurman}, J.~B., {et~al.} 1997, \solphys,
  175, 393

\bibitem[{{Del Zanna} {et~al.}(1997){Del Zanna}, {Hood}, \&
  {Longbottom}}]{1997A&A...318..963D}
{Del Zanna}, L., {Hood}, A.~W., \& {Longbottom}, A.~W. 1997, \aap, 318, 963

\bibitem[{{Doyle} {et~al.}(1998){Doyle}, {Banerjee}, \&
  {Perez}}]{1998SoPh..181...91D}
{Doyle}, J.~G., {Banerjee}, D., \& {Perez}, M.~E. 1998, \solphys, 181, 91

\bibitem[{{Giordano} {et~al.}(2000){Giordano}, {Antonucci}, {Noci}, {Romoli},
  \& {Kohl}}]{2000ApJ...531L..79G}
{Giordano}, S., {Antonucci}, E., {Noci}, G., {Romoli}, M., \& {Kohl}, J.~L.
  2000, \apjl, 531, L79

\bibitem[{{Golub} \& {Pasachoff}(2009)}]{2009soco.book.....G}
{Golub}, L. \& {Pasachoff}, J.~M. 2009, {The Solar Corona} (Cambridge
  University Press, Cambridge)

\bibitem[{{Harra} {et~al.}(2015){Harra}, {Baker}, {Edwards}, {Hara}, {Howe}, \&
  {van Driel-Gesztelyi}}]{2015SoPh..tmp....8H}
{Harra}, L., {Baker}, D., {Edwards}, S.~J., {et~al.} 2015, \solphys, 000, 000

\bibitem[{{Heyvaerts} \& {Priest}(1983)}]{1983A&A...117..220H}
{Heyvaerts}, J. \& {Priest}, E.~R. 1983, \aap, 117, 220

\bibitem[{{Hood} {et~al.}(2002){Hood}, {Brooks}, \&
  {Wright}}]{2002RSPSA.458.2307W}
{Hood}, A.~W., {Brooks}, S.~J., \& {Wright}, A.~N. 2002, Royal Society of
  London Proceedings Series A, 458, 2307

\bibitem[{{Jess} {et~al.}(2012){Jess}, {De Moortel}, {Mathioudakis},
  {Christian}, {Reardon}, {Keys}, \& {Keenan}}]{2012ApJ...757..160J}
{Jess}, D.~B., {De Moortel}, I., {Mathioudakis}, M., {et~al.} 2012, \apj, 757,
  160

\bibitem[{{Malara} {et~al.}(2000){Malara}, {Petkaki}, \&
  {Veltri}}]{2000ApJ...533..523M}
{Malara}, F., {Petkaki}, P., \& {Veltri}, P. 2000, \apj, 533, 523

\bibitem[{{Matteini} {et~al.}(2015){Matteini}, {Horbury}, {Pantellini},
  {Velli}, \& {Schwartz}}]{2015ApJ...802...11M}
{Matteini}, L., {Horbury}, T.~S., {Pantellini}, F., {Velli}, M., \& {Schwartz},
  S.~J. 2015, \apj, 802, 11

\bibitem[{{Nakariakov} {et~al.}(1998){Nakariakov}, {Roberts}, \&
  {Murawski}}]{1998A&A...332..795N}
{Nakariakov}, V.~M., {Roberts}, B., \& {Murawski}, K. 1998, \aap, 332, 795

\bibitem[{{Ofman} \& {Davila}(1997)}]{1997ApJ...476..357O}
{Ofman}, L. \& {Davila}, J.~M. 1997, \apj, 476, 357

\bibitem[{{Raouafi} {et~al.}(2007){Raouafi}, {Harvey}, \&
  {Solanki}}]{2007ApJ...658..643R}
{Raouafi}, N.-E., {Harvey}, J.~W., \& {Solanki}, S.~K. 2007, \apj, 658, 643

\bibitem[{{Similon} \& {Sudan}(1989)}]{1989ApJ...336..442S}
{Similon}, P.~L. \& {Sudan}, R.~N. 1989, \apj, 336, 442

\bibitem[{{Smith} {et~al.}(2007){Smith}, {Tsiklauri}, \&
  {Ruderman}}]{2007A&A...475.1111S}
{Smith}, P.~D., {Tsiklauri}, D., \& {Ruderman}, M.~S. 2007, \aap, 475, 1111

\bibitem[{{Tsiklauri}(2014)}]{2014PhPl...21e2902T}
{Tsiklauri}, D. 2014, Physics of Plasmas, 21, 052902

\bibitem[{{Tsiklauri} {et~al.}(2003){Tsiklauri}, {Nakariakov}, \&
  {Rowlands}}]{2003A&A...400.1051T}
{Tsiklauri}, D., {Nakariakov}, V.~M., \& {Rowlands}, G. 2003, \aap, 400, 1051

\bibitem[{{Tu} \& {Marsch}(1994)}]{1994JGR....9921481T}
{Tu}, C.-Y. \& {Marsch}, E. 1994, \jgr, 99, 21481

\bibitem[{{Young}(2015)}]{2015ApJ...801..124Y}
{Young}, P.~R. 2015, \apj, 801, 124

\end{thebibliography}

\end{document}